\documentclass[11pt,a4paper]{amsart}
\usepackage{placeins}

\usepackage{amssymb}
\usepackage{amsthm, amsmath, amsfonts, amssymb}
\usepackage{array}
\usepackage{mathtools}
\usepackage{algorithm}
\usepackage{algpseudocode}
\usepackage{graphicx}
\usepackage{textcomp}
\usepackage{subfig}  
\usepackage[dvipsnames]{xcolor}
\usepackage{float}
\usepackage{xprintlen}
\usepackage{colortbl}
\usepackage{gensymb}
\usepackage{multirow}
\usepackage{marginnote}
\usepackage{appendix}
\usepackage[T1]{fontenc}
\usepackage{soul}
\usepackage[]{pdfcomment}

\theoremstyle{definition}

\newcommand{\R}{\mathbb R}

\newcommand{\Dc}{\mathcal{D}}

\usepackage[symbol]{footmisc}

\newcommand{\kdeim}{k_{\text{D}}}
\newcommand{\ktay}{k_{\text{T}}}
\DeclareMathOperator{\diag}{diag}

\begin{document}
\title{Parameter estimation and model reduction for retinal laser treatment}
\author[M.\ Schaller et al.]{Manuel Schaller$^{1}$, Mitsuru Wilson$^{1}$, Viktoria Kleyman$^{2}$, Mario~Mordmüller$^{3}$, Ralf~Brinkmann$^{3,4}$, Matthias A.\ Müller$^{2}$ and Karl~Worthmann$^{1}$}
	\thanks{}

\thanks{$^{1}$Technische Universit\"at Ilmemau, Institute of Mathematics, Germany (e-mail: \{manuel.schaller,mitsuru.wilson,karl.worthmann\}@tu-ilmenau.de).}
\thanks{$^{2}$Leibniz University Hannover, Institute of Automatic Control, Germany (e-mail: \{kleyman,mueller\}@irt.uni-hannover.de).}
\thanks{$^{3}$University of Lübeck, Institute of Biomedical Optics, Germany (e-mail: \{ralf.brinkmann,m.mordmueller\}@uni-luebeck.de).}

	\thanks{{\bf Acknowledgments: }The collaborative project "Temperature controlled retinal laser treatment" is funded by the German Research Foundation (DFG) under the project number 430154635 (MU 3929/3-1, WO 2056/7-1, BR 1349/6-1). MS was also funded by the DFG (grant WO\ 2056/2-1, project number 289034702). KW gratefully acknowledges funding by the German Research Foundation (DFG; grant WO\ 2056/6-1, project number 406141926). }.

\begin{abstract}% Abstract of not more than 250 words.
\noindent	Laser photocoagulation is one of the most frequently used treatment approaches for retinal diseases such as diabetic retinopathy and macular edema. The use of model-based control, such as Model Predictive Control (MPC), enhances a safe and effective treatment by guaranteeing temperature bounds. In general, real-time requirements for model-based control designs are not met since the temperature distribution in the eye fundus is governed by a heat equation with a nonlinear parameter dependency. This issue is circumvented by representing the model by a lower-dimensional system which well-approximates the original model, including the parametric dependency. We combine a global-basis approach with the discrete empirical interpolation method, tailor its hyperparameters to laser photocoagulation, and show its superiority in comparison to a recently proposed method based on Taylor-series approximation. Its effectiveness is measured in computation time for MPC. 
We further present a case study to estimate the range of absorption parameters in porcine eyes, and by means of a theoretical and numerical sensitivity analysis we show that the sensitivity of the temperature increase is higher with respect to the absorption coefficient of the retinal pigment epithelium (RPE) than of the choroid's. 

\end{abstract}

\maketitle
\section{Introduction}
\label{lsec:intro}
\noindent 
	Laser photocoagulation is a treatment for a variety of retinal diseases. Recently a non-damaging thermal stimulation of the retina is becoming more and more popular. However, in this case, the irradiated areas on the retina are invisible and proper dosing becomes a challenge, in contrast to standard photocoagulation with visible spots. Due to the strongly varying absorption at the retina, a constant laser power leads to significantly different temperature increases and makes a defined and safe hyperthermia practically impossible.

\begin{figure}[h]
	\centering
	\includegraphics[width =.8\linewidth]{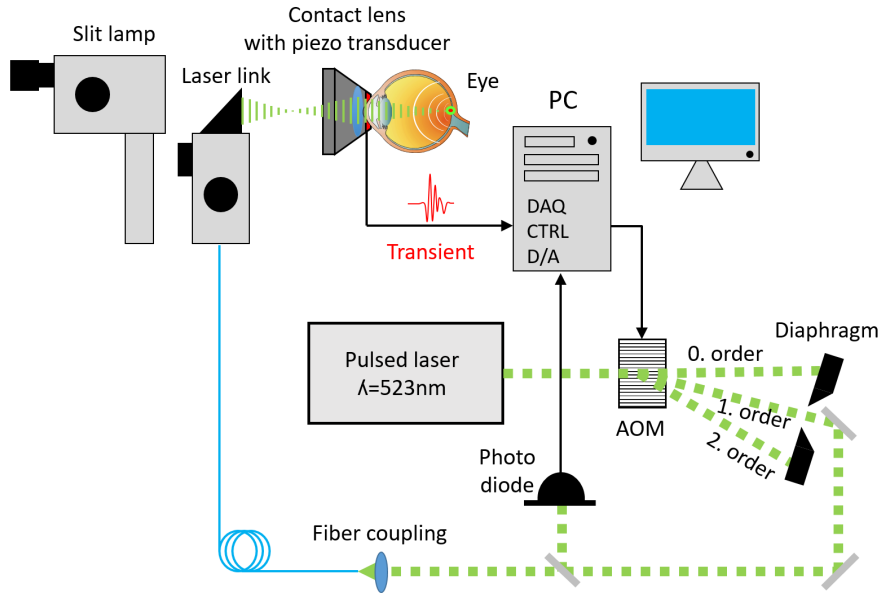}
	\caption{Schematic sketch of the experimental setup.}
	\label{fig:Setup_01}
\end{figure}

	A non-invasive method for determining an average depth-weighted volume temperature based on pressure wave measurements was developed in \cite{Brinkmann.2012}. This real-time temperature feedback allows for the development of controls to obtain homogeneous treatment results, independent of the absorption inside the tissue. 
	A sketch of the experimental setup is depicted in Fig.~\ref{fig:Setup_01}. The beam of a pulsed, solid state Nd:YLF laser with a wave length of $523\, \text{nm}$ is coupled to an optical fiber and guided to the slit lamp. After leaving the slit lamp, the beam is focused onto the tissue sample by means of an ophthalmic contact lens. The contact lens was customized with a ring shaped piezo-ceramic transducer as a pressure sensor and attached to a sample cuvette. The amplitude of the pressure wave is then used to calculate the volume temperature. The laser beam is aimed through an acousto-optic modulator (AOM). Upon HF-modulation of the AOM, the laser beam is split into different orders of diffraction. Here, only the first order of diffraction is used for sample irradiation. Both, pressure transient and laser pulse signals are recorded by a fast data acquisition board and processed with C/C++ MFC software. This enables to normalize the pressure transients to the laser pulse energy and hence to compensate for laser pulse energy fluctuations.

	The laser is operated with a pulse repetition rate of $10 \; \text{kHz}$. Every 10th pulse is set to a fixed probe energy and used for temperature measurement. All experiments are conducted on retinal pigment epithelium (RPE) explants of enucleated porcine eyes with removed retina. The RPE is the major absorber in the eye. For more details on the experimental setup and measurement routines, we refer to \cite{Mordmueller2021}.

	The volume temperature that can be obtained as a measurement from the setup depicted in Figure~\ref{fig:Setup_01} can only be used as a control variable to a limited extent, since the peak temperature within the tissue affects the treatment outcome significantly. Therefore, an estimation of the peak temperature is required, which is achieved via a model-based approach. To this end, we modeled the heat diffusion within the tissue using finite differences in \cite{Kleyman2020a}. However, the dimension of the spatially discretized model (>80000) is computationally intractable for real-time estimation/control algorithms in the kHz-range. Thus, we presented in \cite{Kleyman2020b} an approach adapting the parametric model reduction method from \cite{Baur11} to obtain a low-dimensional representation of the heat equation which retains the absorption dependence and is suitable for applications in the kHz-range. Further, we presented first results regarding observers for the reduced models states and the absorption coefficient, such as an extended Kalman filter and a moving horizon estimator in \cite{Kleyman2021}. 	In~\cite{Mordmueller2021}, we proposed a Model Predictive Control (MPC) scheme %, an optimization-based model-based control technique,
to ensure a safe and effective treatment. As a core feature, MPC enables us to directly formulate bounds on the peak temperature in the underlying optimization problem that is used to evaluate the feedback controller.

	In this paper, we thoroughly compare two parametric model reduction techniques in our particular application for the consideration of either one or two independent absorption coefficients in the eye. First, we extend the approach of \cite{Kleyman2020b,Kleyman2021} to the case of two parameters, where we combine a Taylor series expansion of the nonlinear parameter dependency and with the interpolation based parametric model order reduction method (pMOR) from \cite{Baur11}. Second, we consider a global basis (gb) approach \cite[Section 4.1]{benner2015survey}, where we construct a basis from system snapshots sampled at different parameters. In order to be able to evaluate the nonlinear parameter dependency efficiently, we pair this approach with a discrete empirical interpolation method (DEIM) \cite{Chaturantabut2010}. Both, the presented pMOR and global basis approach utilize an Iterative Rational Krylov Algorithm (IRKA \cite{GugeAnto08}) that is particularly well-suited for sparse and large matrices resulting from discretization of the heat diffusion equation. While both methods yield a small approximation error in terms of the volume and the peak temperature for small projection orders, the DEIM+gb yields superior performance w.r.t.\ the approximation error in terms of both measures --~independently of the used norm (maximum or $L_2$-error). We further illustrate, that the obtained reduced models enable model predictive control at a one\;kHz sampling rate.

	To perform either of the two presented parametric model reduction techniques, a good understanding of the parameter domain is necessary. Thus, we conduct an extensive case study with porcine eyes to estimate the absorption coefficients for different eyes and spots. This case study provides us information about the parameter distribution, i.e., the mean and the variance of the absorption coefficients. Further, we provide confidence intervals for the estimated absorption coefficients and analyze theoretically and numerically the sensitivity of the model w.r.t.\ changes in either parameter, i.e., the absorption coefficient of the RPE and the choroid. We show that the sensitivity regarding the absorption coefficient of the choroid is significantly smaller than its counterpart w.r.t.\ the RPE. Further, we carry out numerical experiments to estimate the effect of fixing the absorption coefficient in the choroid and only estimating the absorption coefficient in the RPE. We display that when taking into consideration other error sources, such as modeling, discretization, model order reduction and measurement noise, this error is relatively small. Hence, in view of the real-time requirements of our application, this could serve as a basis to only estimate the dominant RPE absorption coefficient in order to further reduce the computation times.

	This work is organized as follows. In Section~\ref{sec:mod_disc}, we introduce the PDE-model describing the heat absorption in the eye induced by laser treatment and the corresponding space-time discretization. In Section~\ref{sec:case}, we present a case study for absorption coefficients in the RPE and the choroid for porcine eyes. After describing the methodology, we present results for 250 treatment spots and investigate, e.g., confidence intervals or the spatial distribution of absorption over one explant. In Section~\ref{sec:sensi}, we provide a sensitivity analysis of the input and output map of our model with respect to the parameters theoretically and numerically in time and frequency domain. In Section~\ref{sec:MOR}, we compare two parametric model reduction techniques for both the case of one absorption parameter and the case of two absorption parameters. The real-time capability of an MPC approach using this reduced model is shown in Section~\ref{sec:mpc}. Finally, we conclude and give an outlook regarding future work.

\section{Modeling and discretization}
\label{sec:mod_disc}
\begin{figure}[ht]
	\centering  
	\includegraphics[width =.6\linewidth]{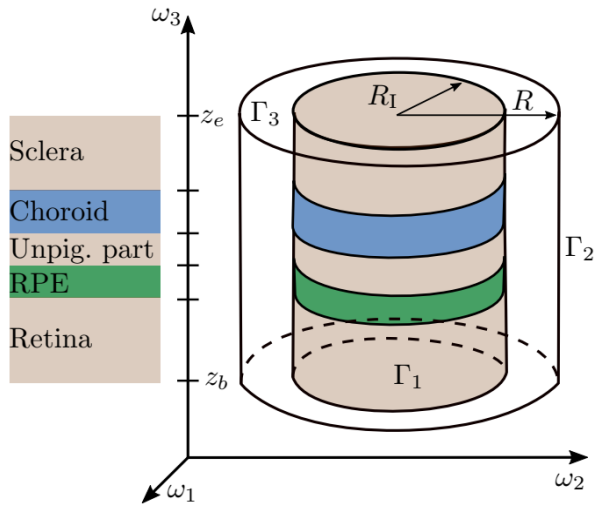}
	\caption{Schematic illustration of the five considered layers of the eye fundus and the cylinders. Figure adapted from \cite{Kleyman2020b}.}
	\label{fig:cylinders}
\end{figure}

\noindent In this section, we briefly describe the PDE-model for heat absorption used for retinal laser treatment as well as its time and space discretization. For more details, we refer to the previous works \cite{Kleyman2020b,Kleyman2021}. The computational domain, denoted by $\Omega \subset \mathbb{R}^3$, is depicted in Figure~\ref{fig:cylinders} and consists of a cylinder with radius $R$ that encloses the irradiated area, given by a smaller cylinder with radius $R_\text{I}$. %
We consider five different layers in the eye fundus, where, however absorption only takes place in the choroid and the retinal pigment epithelium (RPE). The radius $R$ of the larger cylinder is chosen large enough such that we can safely assume that the temperature change during treatment is close to zero at its boundary, allowing us to set homogeneous Dirichlet boundary conditions in the PDE model. The boundary of our spatial domain will be denoted by $\Gamma~=~\Gamma_1\cup\Gamma_2\cup\Gamma_3$.
\medskip

\noindent\textbf{Modeling.}
We will denote by $x(t,\omega)$ the temperature difference with respect to the ambient temperature at time instance $t\in \mathbb{R}^+$ and space point $\omega=(\omega_1,\omega_2,\omega_3)\in \Omega$. Incorporating the heat source resulting from the laser power profile $u : \mathbb{R}^+ \to \mathbb{R}$ by means of the Lambert-Beer law, the evolution of the temperature distribution can be described by the linear parabolic PDE
\begin{align}
    \rho C_\text{p} \frac{\partial x(t,\omega)}{\partial t}-k\Delta x(t,\omega)=u(t)\frac{\chi_{R_\text{I}}(\omega)}{\pi R_\text{I}^2}\mu(\omega_3) e^{-\int_0^{\omega_3}\mu(\zeta)\text{d}\zeta}
    \label{eq:pde}
\end{align}
for all $(t,\omega)\in\mathbb{R}^+\times \Omega$, where $\chi_{R_I}$ is the characteristic function of the interior cylinder in Figure~\ref{fig:cylinders}, i.e., $\chi_{R_\text{i}}(\omega) = 1$ if $\omega_1^2+\omega_2^2 \leq R_\text{I}^2$ and zero otherwise. The symbol $\Delta=\frac{\partial^2}{\partial^2\omega_1}+\frac{\partial^2}{\partial^2\omega_2}+\frac{\partial^2}{\partial^2\omega_3}$ denotes the Laplace operator. The boundary and initial conditions are given by
\begin{align}
\begin{split}
x(t,\omega)&=0\quad \text{$\forall\, (t,\omega)\in\mathbb{R}^+\times \Gamma$},\\ 
x(0,\omega)&=0 \quad \forall\, \omega\in\Omega.
\end{split}
\label{eq:BC}
\end{align}
The heat capacity $C_\text{p}$, the thermal conductivity $k$ and the density $\rho$ are assumed to be constant and the same to those of water (${\rho = 993\text{ kg/m}^3}$, $C_\text{p} = 4176\text{ J/(kgK)}$, ${k = 0.627\text{ W/mK}}$), the main component of tissue, cf.~\cite{Baade.2017}.

The absorption is governed by the scalar valued function $\mu:\Omega \to \mathbb{R}^+$ that is defined piecewise via
\begin{align*}
\mu(\omega_3) = \begin{cases}
\mu_\text{RPE}, \quad &\text{if } \omega_3 \in \text{RPE},\\
\mu_\text{ch}, \quad &\text{if } \omega_3 \in \text{choroid},\\
0, &\text{otherwise}.
\end{cases}
\end{align*}
 In Table~\ref{tab:absorption}, we provide the reference values we used for the thicknesses, cf.~Figure~\ref{fig:cylinders}, and the absorption coefficients.
\begin{table}[ht]
	\centering
	\begin{tabular}{| l | |c|c|}
	\hline
		 & Thickness ($10^{-6}\,\text{m}$) & Absorp.\ coeff.\ ($10^2\,\text{m}^{-1}$)\\
		\hline
		Sclera & $d_\text{sc} = 139\,\,$ & $0$\\
		Choroid & $d_\text{ch}=400\,\,$ &$\mu_\text{ch}^0 =270\phantom{4}$\\
		Unpig.\ & $d_\text{up}=4\quad\,\,\,$ & $0$ \\
		RPE & $\,d_\text{rpe} =6\quad\,\,\,\,\,$ & $\mu_\text{RPE}^0 = 1204$ \\
		Retina & $\,d_\text{r} = 190\,$ & $0$\\
		\hline
	\end{tabular}
		\caption{Average thicknesses and absorption coefficients in porcine eyes from \cite{Brinkmann.2012}}
	\label{tab:absorption}
\end{table}

\noindent The PDE \eqref{eq:pde} can be restated as an abstract infinite-dimensional control system
\begin{align}
\label{eq:abstractevo}
    \dot x(t) = \mathcal{A}x(t) + \mathcal{B}(\mu)u(t)\qquad x(0)=0
\end{align}
that is governed by an unbounded operator $\mathcal{A}:D(\mathcal{A})\subset L_2(\Omega)\to L_2(\Omega)$, and an input operator $\mathcal{B}(\mu)\in L(\mathbb{R},L_2(\Omega))$. Here, $\mathcal{A} = \frac{k}{\rho C_\text{p}}\Delta,$ $D(\mathcal{A}) = H^1_0(\Omega)\cap H^2(\Omega)$ and
\begin{align}
\label{eq:input_operator}
\mathcal{B}(\mu) = \frac{\chi_{R_\text{I}}(\omega)}{\rho C_\text{p}\pi R_\text{I}^2}\mu(\omega_3) e^{-\int_0^{\omega_3}\mu(\zeta)\text{d}\zeta}.
\end{align}
Well-posedness of this system is guaranteed by classical semigroup theory, cf.\ \cite[Section 2]{Curtain1995} and we will omit the details here. We will consider two output relations in our application. The first output is given by a volume temperature and represents the quantity that we can measure by means of the piezo transducer, cf.\ Figure~\ref{fig:Setup_01}. The corresponding output operator $\mathcal{C}(\mu)\in L(L_2(\Omega),\mathbb{R})$ is, using cylinder coordinates and rotational symmetry \cite[Section 2.2]{Kleyman2020b}, given by
\begin{align}
\label{eq:cvol}
\mathcal{C}_\text{vol}(\mu)x &= \int_{z_\text{b}}^{z_\text{e}} x_\text{mean}(t,\omega_3) \mu(\omega_3)e^{\int_{0}^{\omega_3}\mu(\zeta)\text{d} \zeta}\, \text{d}\omega_3,\\
\intertext{where} 
\label{eq:mean}
x_\text{mean}(t,\omega_3) &=\frac{1}{\pi R_\text{I}^2}\int_{0}^{2\pi} \text{d}\phi \int_{0}^{R_\text{I}} r x(r,\omega_3,t)\,\text{d}r.
\end{align}
The second output relation that models the peak temperature in the tissue directly corresponds to success of the treatment and hence is particularly important for control. The corresponding output operator $\mathcal{C}_\text{peak}:C(\Omega)\to \mathbb{R}$ is defined by the temperature at the center of the RPE, i.e., in cartesian coordinates,
\begin{align}
\label{eq:cpeak}
\mathcal{C}_\text{peak}x = x\left(0,0,z_b + d_\text{Retina}+\frac{d_\text{RPE}}{2}\right).
\end{align}
This operator serves as a linear and differentiable approximation of the maximal temperature $\max_{\omega \in \Omega} x(\omega)$. This approximation is justified by numerical experiments, which showed that during heating and up to a very short initial phase, the maximal temperature is attained at the center of the RPE.\\
\noindent\textbf{Discretization.} Applying the finite-difference method to the reformulation of \eqref{eq:pde} using cylindrical coordinates (cf.~\cite{Kleyman2020b}) gives a finite dimensional state space model,
 \begin{align}
 \nonumber
	\dot{x}(t)&= A_cx(t) + B(\mu)u(t)\\
	\label{eq:discreteevo}
	y_\text{vol}(t)&=C_\text{vol}(\mu)x(t)\\
	\nonumber
	y_\text{peak}(t)&=C_\text{peak}x(t)
\end{align}
with $n\in \mathbb{N}$ large, $A_c\in \mathbb{R}^{n\times n}$ and $B,C_\text{vol}:\mathbb{R}\to \mathbb{R}^n$ and $C_\text{peak}\in \mathbb{R}^n$. In order to resolve also the thin layers with a uniform discretization, cf.\ Table~\ref{tab:absorption}, we obtain a high-dimensional model with $n>80000$ degrees of freedom.

For a fixed time step size $\delta>0$, and for $A:= (I - \delta A_c)^{-1}$, the implicit Euler method yields the following discrete system:
\begin{align}
	\label{eq:discretemodel}
	\begin{split}
		x_{k+1}& = A(x_k + \delta B(\mu)u_k)\\
		y_{\text{vol},k} &= C_\text{vol}(\mu)x_k\\
		y_{\text{peak},k} &= C_\text{peak}x_k.
	\end{split}
\end{align}

\section{Absorption coefficients in porcine eyes: A case study}
\label{sec:case}
\noindent 
As can be observed in experiments, the absorption coefficients $\mu_\text{RPE}$ and $\mu_\text{ch}$ that enter the input and output operator in a nonlinear exponential fashion are highly spot and patient dependent. In particular, they can significantly deviate from the reference values given in Table~\ref{tab:absorption}, which necessitates online parameter estimation in treatment. Hence, we compare suitable MOR approaches that retain the parametric dependency (pMOR) in the second part of this paper. However, in order to apply these methods it is first necessary to have information about the range of parameters that can occur and for which the parametric reduced order surrogate model needs to be valid.

	To this end, in this section we carry out a case study of the absorption coefficients of porcine eyes appearing in the input and output operator in the PDE-model \eqref{eq:pde} resp.\ its fully discretized counterpart \eqref{eq:discretemodel}. We conducted experiments on 250 treatment spots and we subsequently used a least square parameter estimation to identify the absorption coefficients at each of these spots. The identified absorption coefficients vary greatly from spot to spot and the domain of the identified absorption coefficients was found to be wide. The identification of the range of absorption coefficients is crucial for the parametric model order reduction techniques in Section~\ref{sec:MOR}. 

	After presenting the methodology in Subsection~\ref{subsec:case:method}, we present in Subsection~\ref{subsec:case:results} the results of the case study, such as detailed values of the absorption coefficients, corresponding confidence intervals, empirical means and empirical standard deviations. In Subsection~\ref{subsec:case:horizon}, we briefly comment on the dependence of the input signal and the identification horizon length on the quality of parameter estimation. 

To avoid scaling issues in the optimization procedure, we parameterize the absorption coefficients relative to the values in the literature $\mu^0_\text{RPE}$ and $\mu^0_\text{ch}$ as given in Table~\ref{tab:absorption}.. That is, we set
\begin{align}
\label{eq:param}
\mu_\text{RPE}(\alpha)=\alpha_\text{RPE}\mu^0_\text{RPE}\qquad \text{and} \qquad
\mu_\text{ch}(\alpha)=\alpha_\text{ch}\mu^0_\text{ch}
\end{align} 
for suitable prefactors $\alpha_\text{RPE},\alpha_\text{ch}\in \mathbb{R}^+$.
Parameter estimation then reduces to estimating these unitless scalar prefactors. We will abbreviate  $\mu= (\mu_\text{RPE},\mu_\text{ch})$ and $\alpha=(\alpha_\text{RPE},\alpha_\text{ch})$ and with slight abuse of notation we will write $B(\alpha) = B(\mu(\alpha))$ and $C(\alpha) = C(\mu(\alpha))$.

\subsection{Methodology for parameter estimation}
\label{subsec:case:method}
\noindent As introduced in the previous section, after space and time discretization the system describing the evolution of the temperature distribution is given by \eqref{eq:discretemodel}. For a given initial temperature distribution $x_0$, the state $x_k$ for $k \geq 1$ can be computed via $x_k = A^k x_0 + \delta \sum_{i=0}^{k-1} A^{k-i} B(\alpha)u_i.$
In this subsection, we consider measurements of the volume temperature, i.e., the first output of \eqref{eq:discretemodel}, obtained from experiments at 250 treatment spots. We formulate least squares parameter estimation problem for $N$ measurements $(y^\text{m}_0,\ldots,y^\text{m}_{N-1})$ via
\begin{align}
\label{e:paramident}
\min_{\alpha \in \mathbb{R}^q} \| F(\alpha)\|^2_2
\end{align}
with $F:\mathbb{R}^2 \to \mathbb{R}^N$ is defined by $F_i(\alpha) = y^\text{m}_{i-1} - C_\text{vol}(\alpha)x_{i-1}$, $i=1,\ldots,N$. As the optimization problem \eqref{e:paramident} is nonlinear and not necessarily convex, we will always refer to local solutions in the following.

\noindent \textbf{Confidence intervals.} Besides the optimal parameters, we will also compute the corresponding confidence intervals for each spot. For a given probability level $p\in (0,1)$, a confidence interval is the region in the parameter space, in which the unknown parameters are located with probability $p$. This confidence region can be estimated by means of covariance analysis of the optimization problem. We follow the standard approach in the literature, see \cite[Chapter 4.3]{Koerkel2002} or \cite{Bock1978}. We will denote by $J(\alpha)$ the Jacobian matrix of $F(\alpha)$. The covariance matrix $\text{Cov}\in \mathbb{R}^{2\times 2}$ of the parameter estimation problem is given by 
\begin{align}
\label{e:conf}
\text{Cov}(\alpha)  =\left(J(\alpha)^\top J(\alpha)\right)^{-1}\in \mathbb{R}^{2\times 2}.
\end{align}
Consider a local solution $\alpha^* \in \mathbb{R}^2$ of \eqref{e:paramident} and by $\alpha^*_i$ its $i$-th component. The confidence interval of probability $p \in (0,1)$ corresponding to the $i$-th parameter, $1\leq i\leq 2$ can be approximated by
\begin{align*}
\left[\alpha_i^* - \sqrt{\gamma(p)\text{Cov}(\alpha^*)_{ii}},\alpha_i^*+ \sqrt{\gamma(p)\text{Cov}(\alpha^*)_{ii}}\right]
\end{align*}
where $\gamma(p) = \chi^2_{p}(1-p)$ is the quantile of the $\chi^2$-distribution with two degrees of freedom. Thus, the width of the confidence intervals is governed by the diagonal entries of the covariance matrix in the optimal parameter configuration, that is, the inverse of $J(\alpha^*)^\top J(\alpha^*)$. 

\subsection{Results}
\label{subsec:case:results}
\begin{figure}
	\centering
\includegraphics[width=0.6\columnwidth]{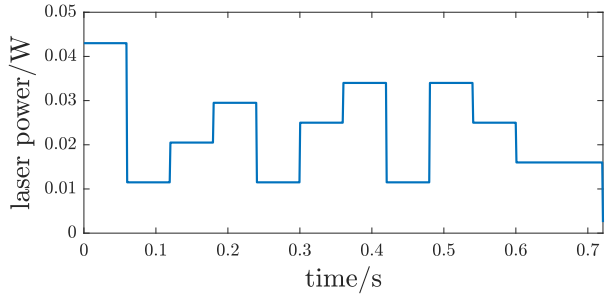}
\caption{Time-varying control used in the numerical and real experiments.}
\label{fig:fancyu}
\end{figure}

\noindent We perform the methodology described in the previous subsection for a total of $n_\text{total} = 250$ measurement spots, $n_\text{spots} =25$ treatment spots in each of the $n_\text{eyes}=10$ eyes. To each spot, we applied a constant laser power of 30\,mW for 720\,ms and after a cooling phase, we applied the time-varying control depicted in Figure~\ref{fig:fancyu}. The volume temperature is measured at the rate of 1kHz and results in $N=721$ measurements per spot for both the constant and the time-varying laser power. The absorption coefficients $\alpha_\text{ch}$ and $\alpha_\text{RPE}$ are then identified as a solution to the optimization problem \eqref{e:paramident} using the measured data.

\begin{figure}[ht]
	\centering
	 \hspace*{-.8mm}\includegraphics[width=0.8\columnwidth]{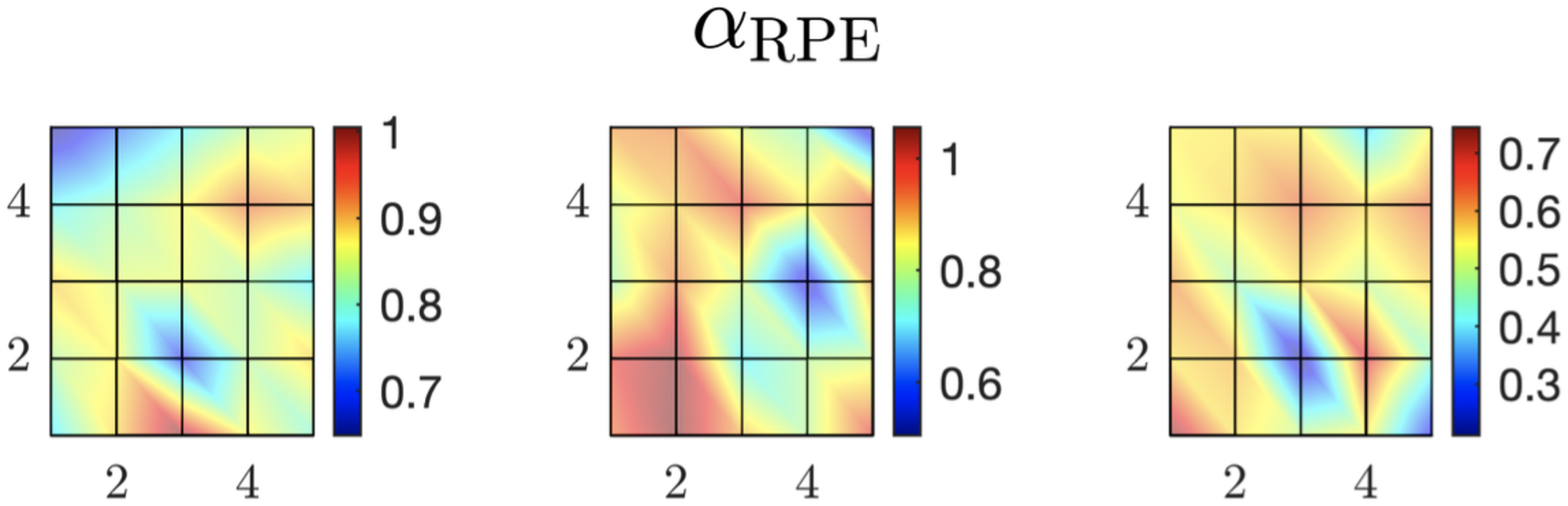}\\
	\includegraphics[width=0.8\columnwidth]{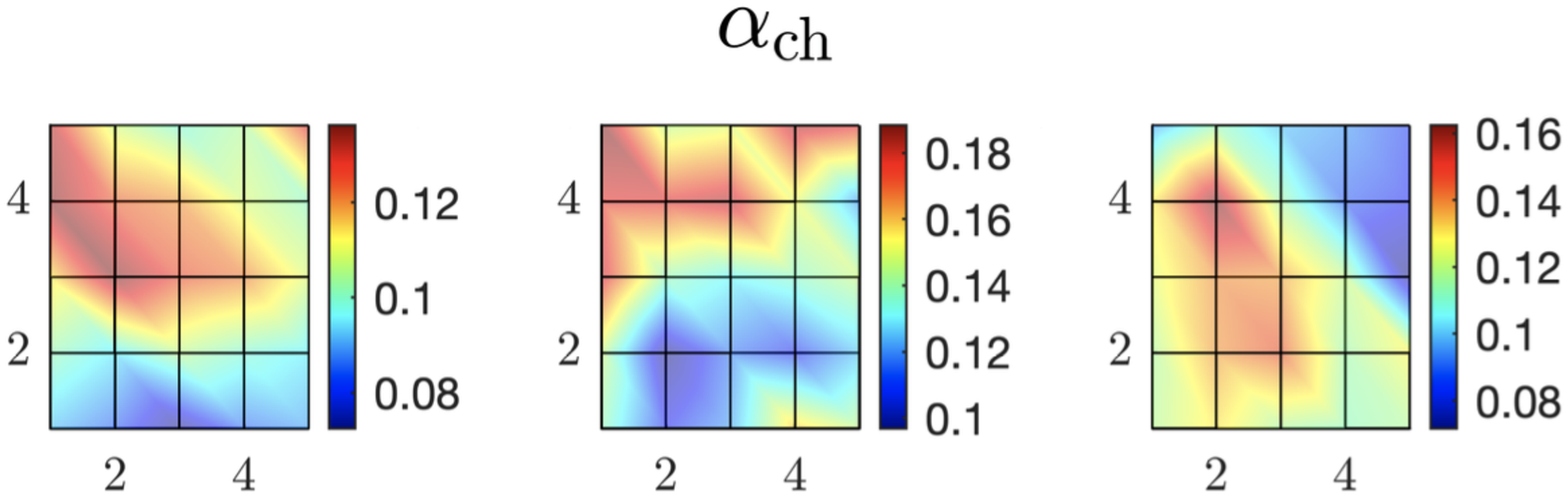}
	\caption{Identified choroid absorption coefficients $\alpha_\text{RPE}$ and $\alpha_\text{ch}$ for eye number 4, 7 and 10 (left to right).}
\label{fig:2dplots1}
\end{figure}

\begin{figure}[ht]
	\includegraphics[width=0.48\columnwidth]{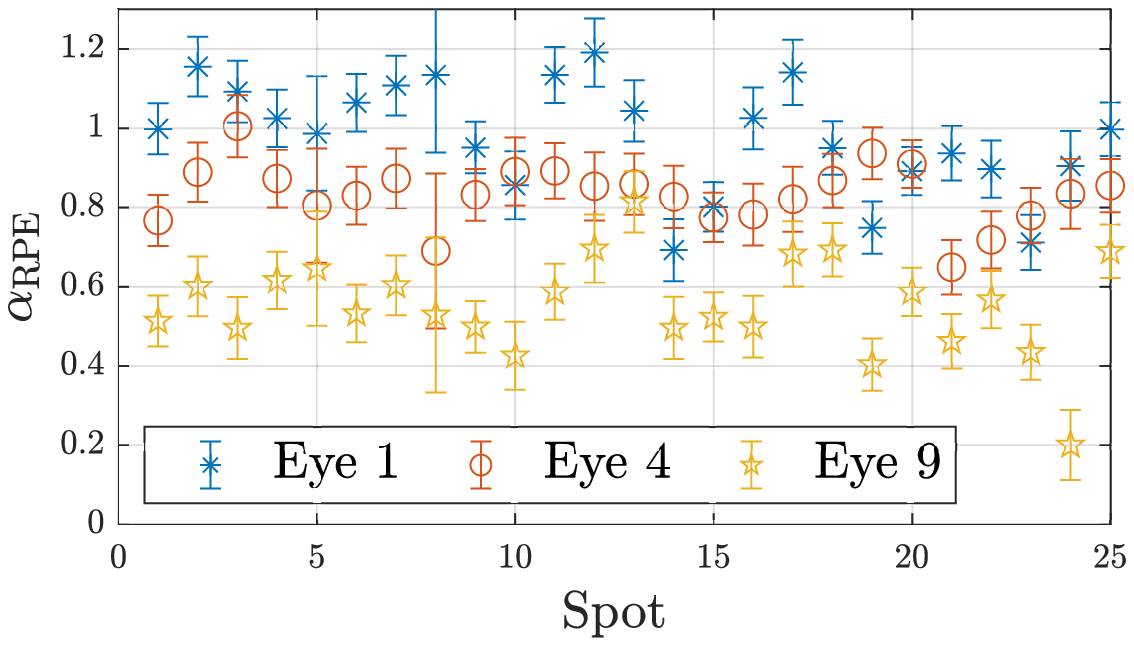}
	\includegraphics[width=0.48\columnwidth]{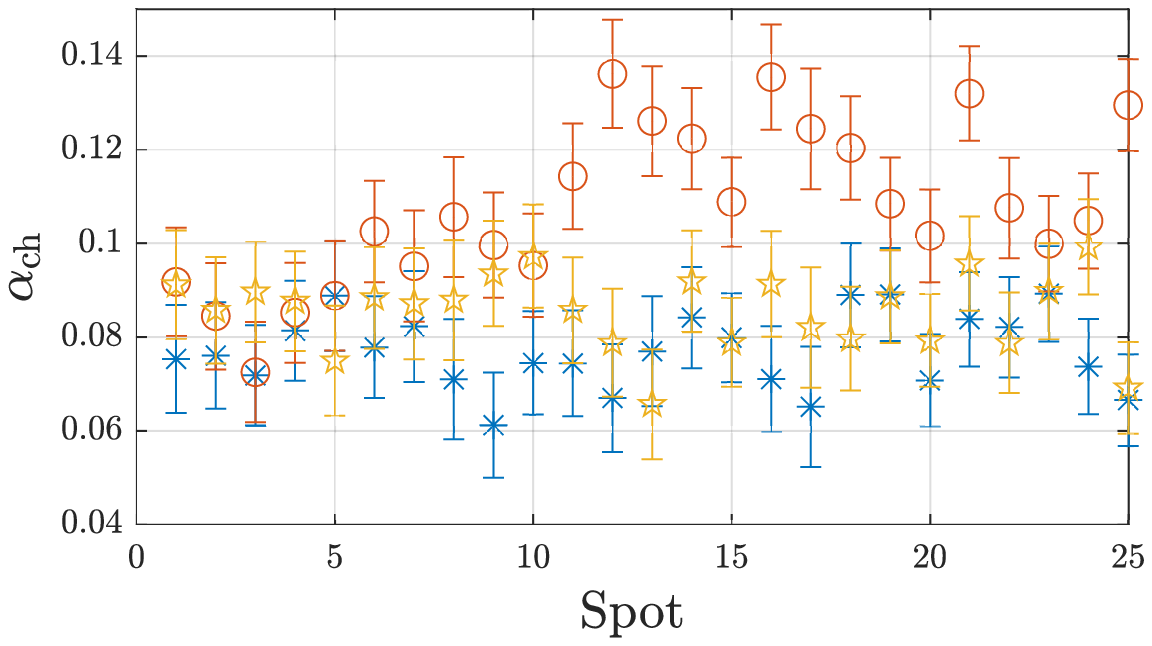}\\

	\caption{Identified absorption coefficients of three eyes for all spots with corresponding 95\%-confidence intervals.}
	\label{fig:casestudy2p}
\end{figure}

	We depict detailed results of the case study for three representative eyes in Figure~\ref{fig:2dplots1} (Spatial distribution for Eye 4,7 and 10) and Figure~\ref{fig:casestudy2p} (Confidence intervals for Eye 1,4 and 9) each. In Figure~\ref{fig:2dplots1}, we can see that two neighboring spots do not necessarily have a similar absorption coefficient. Further, there is no obvious relation between the absorption coefficient in the RPE and its counterpart in the choroid. We observe, however, that the values for RPE for eye 7 are close to its nominal value (i.e., $\alpha_\text{RPE}\approx 1$) whereas the identified values for the choroid are far below its nominal value for all eyes, i.e., $\alpha_\text{ch}\approx 0.1$. One reason could be that there is less blood in the choroid due to the preparation process of the explants. %\ralf{passt die Begründung?}

		In Figure~\ref{fig:casestudy2p}, we we depict the 95\%-confidence intervals along with the values of the identified absorption coefficient in the choroid and the RPE. We observe several outliers, i.e., spot 24 of eye 9, where $\alpha_\text{RPE}$ is the lowest, and in contrast, $\alpha_\text{ch}$ is the highest over all spots. This might be an indication that, due to the non-convexity of the least squares optimization problem \eqref{e:paramident}, the depicted values are local minima.

\noindent The mean and variance of the identified parameters over all eyes and spots is given in Table \ref{t:mean_var}. In order to be able to quantify and analyze the fluctuations in the estimated parameters, we further compute the coefficients of variation $c_\text{ch}$ and $c_\text{RPE}$ of both absorption coefficients for every eye, that is defined by the fraction of empirical standard deviation $\sigma_*$ and empirical mean $\bar{\alpha}_*$, i.e., for $*\in \{\text{RPE},\text{ch}\}$, we have $c_* = \tfrac{\sigma_*}{\bar{\alpha}_*}$, where
\begin{align*}
\bar{\alpha}_* = \tfrac{1}{n_\text{total}} \sum_{i=1}^{n_\text{total}} \alpha_*(i),\quad \sigma_* = \sqrt{\tfrac{1}{n_\text{total}-1} \sum_{i=1}^{n_\text{total}}|\alpha_*(i)- \bar{\alpha}_*|^2}.
\end{align*}
 In Table~\ref{t:mean_var} we can see that the mean and the variance of the identified parameters are almost independent of the type of control used for identification. The slight deviation can stem from unmodeled effects occurring predominantly at either of the controls that distort the estimated parameter.
\begin{table}[ht]
	\centering
	\begin{tabular}{|c|c||c|c|c|}
	\hline
		&&mean $\bar{\alpha}_*$& std.\ dev.\ $\sigma_*$ & coeff.\ var.\ $c_*$ \\\hline\hline
		\multirow{2}{*}{$u\equiv 30 $mW}&$\alpha_\text{RPE}$	&   0.7636    &  0.1907& 0.2498\\
		&$\alpha_\text{ch}$												&   0.0986&  0.0281&0.2853 \\\hline
		\multirow{2}{*}{time var.\ $u$}&$\alpha_\text{RPE}$	&   0.7501    &  0.2198& 0.2931\\
		&$\alpha_\text{ch}$												&   0.1031&  0.0278&0.2691\\
		\hline
	\end{tabular}
	\caption{Results of the case study.}
	\label{t:mean_var}
\end{table}
\subsection{Comparison of identification horizons and control type}
\label{subsec:case:horizon}
\noindent Whereas in the previous part we saw that the identification on the long time horizon of 721\,ms does not strongly depend on the type of control used, we now briefly discuss if this also holds true for smaller time horizons. To this end, we compared the relative error for different identification horizons, i.e., $N\in \{100,200,400\}$ with respect to the identified parameter for the full horizon $N=721$. We observed first, that the relative error is decreasing in horizon length and that the relative error is around 10 percent when using only the first 200 of the 721 measurements. However, both controls performed roughly the same in terms of the approximation quality.

\FloatBarrier

%\newpage
\section{Parametric sensitivity analysis}
\label{sec:sensi}
\noindent In Section \ref{sec:case}, we estimated absorption coefficients for the RPE and the choroid. In particular, the right column of Table~\ref{t:mean_var} suggests that their relative variation from spot to spot is of the same magnitude. In the present section, we quantify the influence that small variations of the parameters have on the input-output behavior by means of a sensitivity analysis. Further, we address the effect of setting the parameter corresponding to the smaller sensitivity constant in order to speed up parameter identification in real-time scenarios without substantially compromising accuracy.
\subsection{Sensitivity analysis of the PDE model}
\label{subsec:theosensi}
\noindent 
	In this subsection, we compute the sensitivity of the input and output map with respect to the parameters. Intuitively, it seems clear that due to the exponentially decaying dependency on the spatial variable $\omega_3$, cf.\ \eqref{eq:input_operator}, the sensitivity with respect to the absorption coefficient of the choroid $\alpha_\text{ch}$ is smaller than w.r.t.\ its counterpart in the RPE. We will supply a theoretical reasoning for this intuition. Recall the input and output operators from \eqref{eq:input_operator} and \eqref{eq:cvol}
\begin{align*}
(\mathcal{B}(\mu))(\omega_3) &= \tfrac{\chi_{R_\text{I}}(\omega)}{\rho C_\text{p} \pi R_i^2}\mu(\omega_3) e^{-\int_0^{\omega_3}\mu(\xi)d\xi}\\
\mathcal{C}_\text{vol}(\mu)x &= \int_{z_b}^{z_e} x_\text{mean}(t,\omega_3)\mu(\omega_3)e^{-\int_0^{\omega_3}\mu(\xi)d\xi}\,d\omega_3.
\end{align*}
To abbreviate notation, we define $g:L^\infty(\Omega) \to L^\infty(\Omega)$ pointwise by
\begin{align*}
(g(\mu))(\omega_3)&:= \mu(\omega_3) e^{-\int_0^{\omega_3}\mu(\xi)d\xi}\\&=  \begin{cases}
\mu_\text{RPE} e^{-\omega_3 \mu_\text{RPE}}&\omega_3 \in \text{RPE}\\
\mu_\text{ch} e^{(-d_\text{RPE}\mu_\text{RPE}-(\omega_3-z_\text{ch})\mu_\text{ch})}&\omega_3 \in \text{choroid},
\end{cases}
\end{align*}
where $z_\text{ch}$ denotes the beginning of the choroid in Figure~\ref{fig:cylinders}.
In the following distinction of cases, in order to shorten notation, we will always consider in the first row the case $\omega_3\in \text{RPE}$ and in the second row the case $\omega_3\in \text{choroid}$.
Together with our parameterization $\alpha=(\alpha_\text{RPE},\alpha_\text{ch})$ of the absorption coefficients given in \eqref{eq:param}, this reads
\begin{align*}
(g(\alpha))(\omega_3)=  \begin{cases}
\alpha_\text{RPE}\mu^0_\text{RPE} e^{-\omega_3 \alpha_\text{RPE}\mu^0_\text{RPE}}\\%&\omega_3 \in \text{RPE}\\
\alpha_\text{ch}\mu^0_\text{ch} e^{(-d_\text{RPE}\alpha_\text{RPE}\mu^0_\text{RPE}-(\omega_3-z_\text{ch})\alpha_\text{ch}\mu^0_\text{ch})}.%&\omega_3 \in \text{ch}.
\end{cases}
\end{align*}
Then, we can concisely denote
\begin{align*}
\mathcal{B}(\alpha) &= \tfrac{\chi_{R_\text{I}}(\omega)}{\rho C_\text{p} \pi R_i^2}g(\alpha)\\
\mathcal{C}_\text{vol}(\alpha)x &= \int_{z_b}^{z_e} x_\text{mean}(\omega_3) (g(\alpha))(\omega_3)\,d\omega_3 = \langle g(\alpha),x_\text{mean}\rangle_{L_2(z_b,z_e)}.
\end{align*}
By means of the variation of constants formula, we can compute the sensitivity of the volume and peak temperature with respect to either parameter $\alpha_*$, $*\in \{\text{RPE},\text{choroid}\}$, which, for any time instance $t\geq 0$ is given by
\begin{align*}
     \tfrac{\partial}{\partial \alpha_*}\big(\mathcal{C}_\text{vol}(\alpha)x(t)\big) 
    & = \tfrac{\partial}{\partial \alpha_*}\mathcal{C}_\text{vol}(\alpha)x(t) + \mathcal{C}_\text{vol}(\alpha)\tfrac{\partial}{\partial \alpha_*} x(t) \\
    & = \tfrac{\partial}{\partial \alpha_*}\mathcal{C}_\text{vol}(\alpha)x(t) + \mathcal{C}_\text{vol}(\alpha)\int_0^t e^{(t-s)A_c} \tfrac{\partial}{\partial \alpha_*} \mathcal{B}(\alpha) u(s)\,\text{d}s
\end{align*}
and
\begin{align*}
	\tfrac{\partial}{\partial \alpha_*}\big(\mathcal{C}_\text{peak}(\alpha)x(t)\big) & = \mathcal{C}_\text{peak}\tfrac{\partial}{\partial \alpha_*} x(t)\\
	& = \mathcal{C}_\text{peak}\int_0^t e^{(t-s)A_c} \tfrac{\partial}{\partial \alpha_*} \mathcal{B}(\alpha) u(s)\,\text{d}s
\end{align*}
where the latter follows as the output operator corresponding to the peak temperature is independent on $\alpha$, i.e., $\tfrac{\partial}{\partial \alpha_*} C_\text{peak}=0$.

 The sensitivity of the input map measured in the state space norm $L_2(\Omega)$ is governed by the sensitivity of $g(\alpha)$ measured in the $L_2(z_b,z_e)$-norm:
\begin{align}
\label{e:inputsensi}
\begin{split}
\left\|\int_0^t \right.&\left.e^{(t-s)A_c} \tfrac{\partial}{\partial \alpha_*} \mathcal{B}(\alpha) u(s)\,\text{d}s\right\|_{L_2(\Omega)}\\
&\leq \int_0^t  \left\|e^{(t-s)A_c} \right\|_{L(L_2(\Omega),L_2(\Omega))} \left\|\tfrac{\partial}{\partial \alpha_*} \mathcal{B}(\alpha)\right\|_{L_2(\Omega)} |u(s)|\,\text{d}s\\
&= \int_0^t  \left\|e^{(t-s)A_c} \right\|_{L(L_2(\Omega),L_2(\Omega))} \tfrac{1}{\rho C_\text{p} \pi R_i} \left\|\tfrac{\partial}{\partial \alpha_*} g(\alpha)\right\|_{L_2(z_b,z_e)} |u(s)|\,\text{d}s,
\end{split}
\end{align}
where the last equality holds as $g(\alpha)$ only depends on the third spatial variable, i.e., the depth.

	Correspondingly, we estimate the sensitivities of the output operator $\mathcal{C}_\text{vol}(\alpha)\in L(L_2(\Omega),\mathbb{R})$. To this end, let $v\in L_2(\Omega)$ and compute
\begin{align}
\label{e:outputsensi}
\begin{split}
\left|\tfrac{\partial}{\partial \alpha_*} \mathcal{C}_\text{vol}(\alpha) v\right| = \langle \tfrac{\partial}{\partial \alpha_*} g(\alpha)(\cdot),v_\text{mean}(\cdot)\rangle_{L_2(z_b,z_e)} \leq \left\|\tfrac{\partial}{\partial \alpha_*} g(\alpha)\right\|_{L_2(z_b,z_e)} \|v\|_{L_2(\Omega)},
\end{split}
\end{align}
where $v_\text{mean}(z)$ is the mean at depth $z$ computed over the radial component, cf.\ \eqref{eq:mean}.
	
	Hence, derivatives of $\mathcal{B}(\alpha)$ and $\mathcal{C}_\text{vol}(\alpha)$ with respect to the prefactors $\alpha$ can now be estimated via the partial derivatives of 
\begin{align*}
g(\alpha) = \begin{cases}
\alpha_\text{RPE}\mu^0_\text{RPE}e^{-\omega_3 \alpha_\text{RPE}\mu^0_\text{RPE}}&\omega_3 \in \text{RPE}\\
\alpha_\text{ch}\mu^0_\text{ch} e^{(-d_\text{RPE} \alpha_\text{RPE}\mu_\text{RPE}^0-(\omega_3-z_\text{ch})\alpha_\text{ch}\mu^0_\text{ch})}&\omega_3 \in \text{choroid}.
\end{cases}
\end{align*}
%To this end, we compute
%\begin{align*}
%(\tfrac{\partial}{\partial \alpha_\text{RPE} } g(\alpha))(\omega_3) = \begin{cases}
%\left(\mu^0_\text{RPE} -  \omega_3 \alpha_\text{RPE}\left(\mu_\text{RPE}^0\right)^2\right)e^{-\omega_3 %\alpha_\text{RPE}\mu^0_\text{RPE}}\\%&\omega_3 \in \text{RPE}\\
%-\alpha_\text{ch}\mu_\text{ch}^0 d_\text{RPE}\mu^0_\text{RPE} e^{(-d_\text{RPE}\alpha_\text{RPE}\mu_\text{RPE}^0-(\omega_3-z_\text{ch})\alpha_\text{ch}\mu_\text{ch}^0)}  %&\omega_3 \in \text{Chroroid}
%\end{cases}
%\end{align*}
%\begin{align*}
%&(\tfrac{\partial}{\partial \alpha_\text{ch} } g(\alpha) )(\omega_3) \\&= \begin{cases}
%0\\%&\omega_3 \in \text{RPE}\\
%\left(\mu_\text{ch}^0 -(\omega_3-z_\text{ch})\alpha_\text{ch}\left(\mu_\text{ch}^0\right)^2\right) %e^{(-d_\text{RPE}\mu_\text{RPE}-(\omega_3-z_\text{ch})\mu_\text{ch})} %&\omega_3 \in \text{ch}
%\end{cases}
%\end{align*}
In order to compare the bounds on the input map sensitivity \eqref{e:inputsensi} and the output map sensitivity \eqref{e:outputsensi}, we compare the sensitivities of $g(\alpha)$ in the $L_2(\Omega)$-norm, i.e., for the mean $\bar{\alpha} = (\bar{\alpha}_\text{RPE},\bar{\alpha}_\text{ch})$ for the constant laser power given in Table~\ref{t:mean_var}, we compute,
\begin{align*}
	\left.\left\|\tfrac{\partial}{\partial \alpha_\text{RPE} } g(\alpha)\middle|_{\alpha=\bar{\alpha}}  \right\|_{L_2(\Omega)}\right.\approx 0.1769,
	\quad 
	\left.\left\|\tfrac{\partial}{\partial \alpha_\text{ch} } g(\alpha)\middle|_{\alpha=\bar{\alpha}}  \right\|_{L_2(\Omega)}\right.\approx 0.1254.
\end{align*}
Using Taylor expansion at the mean $\bar{\alpha}$, we have
\begin{align*}
	g(\bar{\alpha} + \delta \alpha)-g(\bar{\alpha}) = \nabla g(\alpha)\delta \alpha + o(|\delta \alpha|).
\end{align*}
Thus, the influence of a perturbation of one standard deviation in each direction, i.e., $\delta \alpha = (\sigma_\text{RPE},0)$ and $\delta \alpha = (0,\sigma_\text{ch})$, is approximately given by $\left\|\tfrac{\partial}{\partial \alpha_\text{RPE} } g(\alpha)\middle\vert_{\alpha=\bar{\alpha}}  \right\|_{L_2(\Omega)}\sigma_\text{RPE}\approx 0.034$ and $\left.\left\|\tfrac{\partial}{\partial \alpha_\text{ch} } g(\alpha)\middle\vert_{\alpha=\bar{\alpha}}  \right\|\right._{L_2(\Omega)}\sigma_\text{ch}\approx 0.0035$, respectively.

Hence, we conclude that the sensitivity of $g(\alpha)$ with respect to $\alpha_\text{RPE}$ is approximately ten times higher than with respect to $\alpha_\text{ch}$ when considering perturbations of one standard deviation each. This directly translates into upper bounds of the input map sensitivity and output map sensitivity via \eqref{e:inputsensi} and~\eqref{e:outputsensi}. However, it is important to note that we only compute and compare upper bounds on the sensitivites. This will be no longer the case in the the next part, where we will compute the sensitivities of the input output behavior directly.

	\subsection{Sensitivity analysis of input-output behavior in frequency and time domain.}
\label{subsec:input_output_sensi}
	\noindent 
	Following the sensitivity analysis of input and output map of the PDE model in time domain, we now analyze the sensitivity of the steady states by analyzing the sensitivity of the transfer function at zero. Moreover, we analyze the sensitivities in time domain by means of numerical experiments. 

	In frequency domain, we will compute the sensitivities of steady states directly by computing the derivatives of the transfer functions corresponding to the full order model~\eqref{eq:discreteevo} at zero, i.e., setting 
\begin{align}
    \label{e:transfer}
    G_\text{vol}(\alpha):= C_\text{vol}(\alpha)A_c^{-1}B(\alpha), \qquad
    G_\text{peak}(\alpha) := C_\text{peak}A_c^{-1}B(\alpha),
\end{align}
we compare for $*\in \{\text{RPE,choroid}\}$
\begin{align*}
    \frac{\partial}{\partial \alpha_*}G_\text{vol}(\alpha) & = \left(\frac{\partial}{\partial \alpha_*}C_\text{vol}(\alpha)\right)A_c^{-1}B(\alpha) +C_\text{vol}(\alpha)A_c^{-1}\left( \frac{\partial}{\partial \alpha_*}B(\alpha)\right) \\
    \frac{\partial}{\partial \alpha_*}G_\text{peak}(\alpha) & = C_\text{peak}A_c^{-1}\left( \frac{\partial}{\partial \alpha_*}B(\alpha)\right).
\end{align*}
Note that the discretization $A_c$ of the Dirichlet Laplacian is always invertible as the underlying dynamics are exponentially stable or, in other words, the largest eigenvalue of the Dirichlet Laplacian is negative.

\begin{figure}[ht]
	\centering
	\includegraphics[scale=.65]{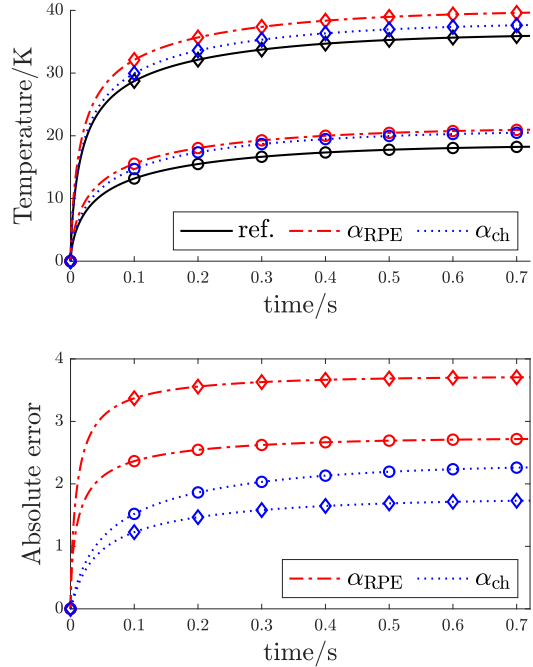}
	\caption{Sensitivities of peak temperature (diamond) and volume temperature (circle) in time domain.}
	\label{fig:sensitivities}
\end{figure}

In Table~\ref{t:transferfunction}, we depict these sensitivities evaluated at the mean of all measurements over all eyes $\bar{\alpha}_*$ as given in Table~\ref{t:mean_var}. We can see, that, when appropriately scaled with one empirical standard deviation, see Table~\ref{t:mean_var}, the sensitivity of volume temperature and peak temperature are higher with respect to the absorption coefficient in the RPE. In the last row of Table~\ref{t:mean_var}, we further scale the values with the constant input 30\,mW in order to compare it later to the time domain sensitivity with respect to the steady state emanating from the constant input 30\,mW in Figure~\ref{fig:sensitivities}.

\begin{table}[ht]
	\centering
		\begin{tabular}{|c||c c c c|} 
		\hline
		&$\tfrac{\partial G_\text{vol}}{\partial \alpha_\text{ch}}$ & $\tfrac{\partial G_\text{vol}}{\partial \alpha_\text{RPE}}$ & $\tfrac{\partial G_\text{peak}}{\partial \alpha_\text{ch}}$ & $\tfrac{\partial G_\text{peak}}{\partial \alpha_\text{RPE}}$ \\ [0.5ex] 
		\hline
		\footnotesize{unscaled}&2.877&0.483&2.934&0.691\\
		\footnotesize{scaled with $\sigma_*$} &0.083&0.092&0.086&0.136\\
		\footnotesize{scaled with $\sigma_*$ and $u \equiv 30\, \text{mW}$}&2.493&2.763&2.565&4.077\\
		\hline
	\end{tabular}
%	\begin{tabular}{|c||c c c c|} 
%		\hline
%		&$\tfrac{\partial G_\text{vol}}{\partial \alpha_\text{ch}}$ & $\tfrac{\partial G_\text{vol}}{\partial \alpha_\text{RPE}}$ & $\tfrac{\partial G_\text{peak}}{\partial \alpha_\text{ch}}$ & $\tfrac{\partial G_\text{peak}}{\partial \alpha_\text{RPE}}$ \\ [0.5ex] 
%		\hline
%		unscaled&2.8767&0.4829&2.9344&0.6911\\
%		scaled with std.\ dev.\ &0.0831&0.0921&0.0855&0.1359\\
%		scaled with std.\ dev.\ and $u \equiv 30 mW$&2.4930&2.7630&2.5650&4.0770\\
%		\hline
%	\end{tabular}
	\caption{Sensitivities of the input-output behavior (W to K) in frequency domain evaluated at the mean of all measurements as given in Table~\ref{t:mean_var}. }
	\label{t:transferfunction}
\end{table}
\noindent In the time domain, we analyze this sensitivity numerically by computing the state along the dynamics of \eqref{eq:discreteevo} for a reference value $\bar{\alpha}$, to which we compare the resulting state trajectories with a perturbation by one standard deviation
\begin{itemize}
\item[a)] of the RPE absorption: $\alpha =  \bar{\alpha} + (\sigma_\text{RPE},0)$ 
\item[b)] of the choroid absorption: $\alpha =  \bar{\alpha} + (0,\sigma_\text{ch})$.
\end{itemize}
The results are given in Figure~\ref{fig:sensitivities}. First, we can see that a one-standard-deviation perturbation of the absorption coefficient induces a perturbation of the volume temperature by roughly 10-30 percent and to a perturbation of the peak temperature by roughly 5-15 percent. Clearly, the influence on both, the volume and the peak temperatures is larger when changing the RPE absorption compared to perturbing the choroid absorption. Finally, we observe that the asymptote of the absolute error in Figure~\ref{fig:sensitivities} is very close to the sensitivities in the third row of Table~\ref{t:transferfunction} for the volume temperature. This can be explained as the steady state $\bar x$ for constant input signal $u \equiv \bar{u}$ is given by $0 = A_c\bar{x}+B(\alpha)\bar{u}$ and hence by invertibility of $A_c$, $\bar{x} = A_c^{-1}B(\alpha)\bar{u}$. Thus, the steady state outputs for $*\in \{\text{vol,peak}\}$ are given by
\begin{align*}
 \bar{y}_* = C_*(\alpha)\bar{x}= C_*(\alpha)A_c^{-1}B(\alpha)\bar{u} = G_*(\alpha)\bar{u},
\end{align*}
where $G_*$, $*\in \{\text{vol,peak}\}$ is defined in \eqref{e:transfer}. The small discrepancy between the asymptotes in Figure~\ref{fig:sensitivities} and the values in Table~\ref{t:transferfunction} could stem from linearization errors as in time domain, we depict the nonlinear sensitivities and in frequency domain we show the sensitivities of first order, i.e., the derivatives. 

\subsection{Only estimating the absorption in the RPE}
\begin{figure}[ht]
	\centering	\hspace*{-.3cm}	\includegraphics[width=.45\columnwidth]{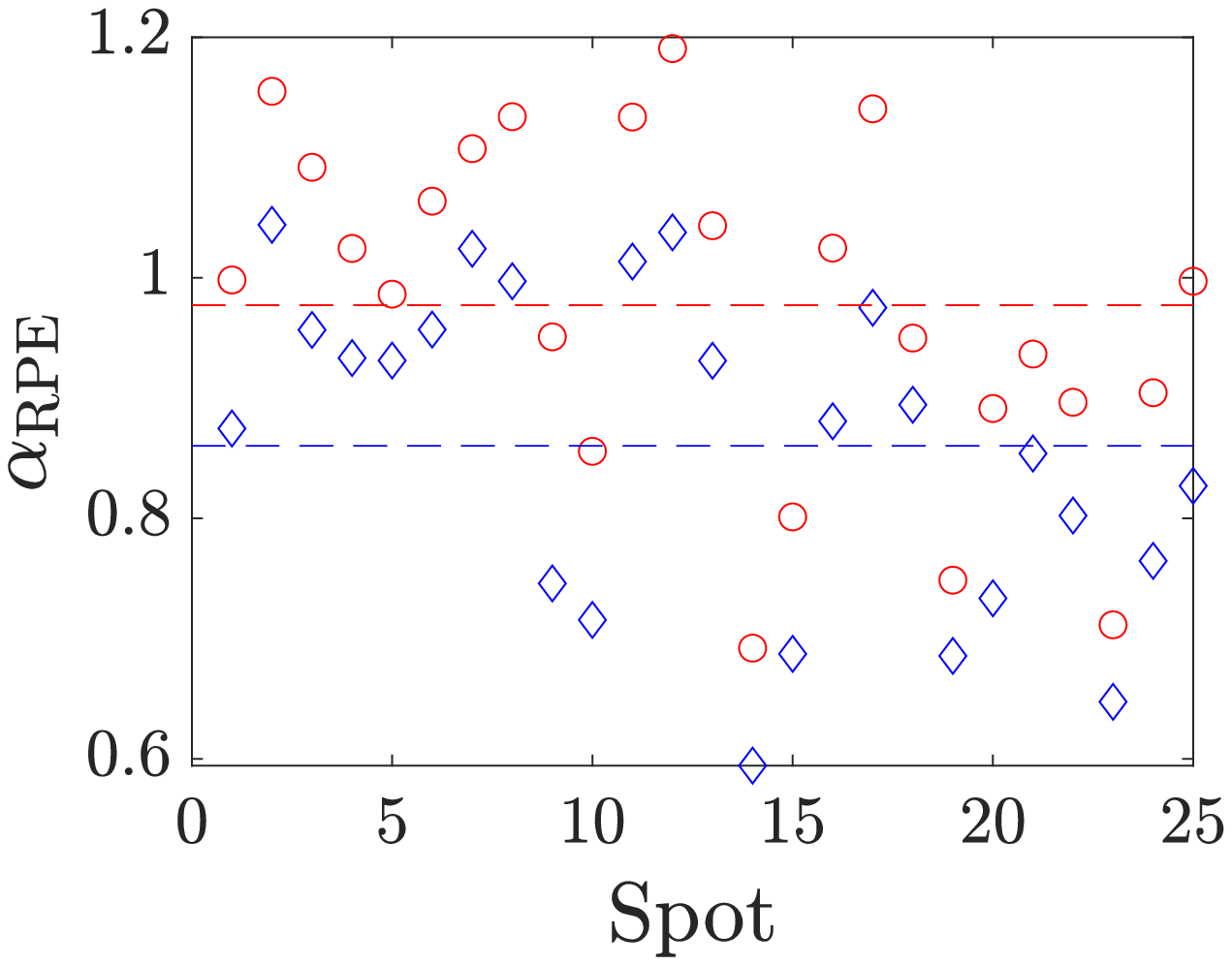}\hspace*{-.03cm}
	\includegraphics[width=.45\columnwidth]{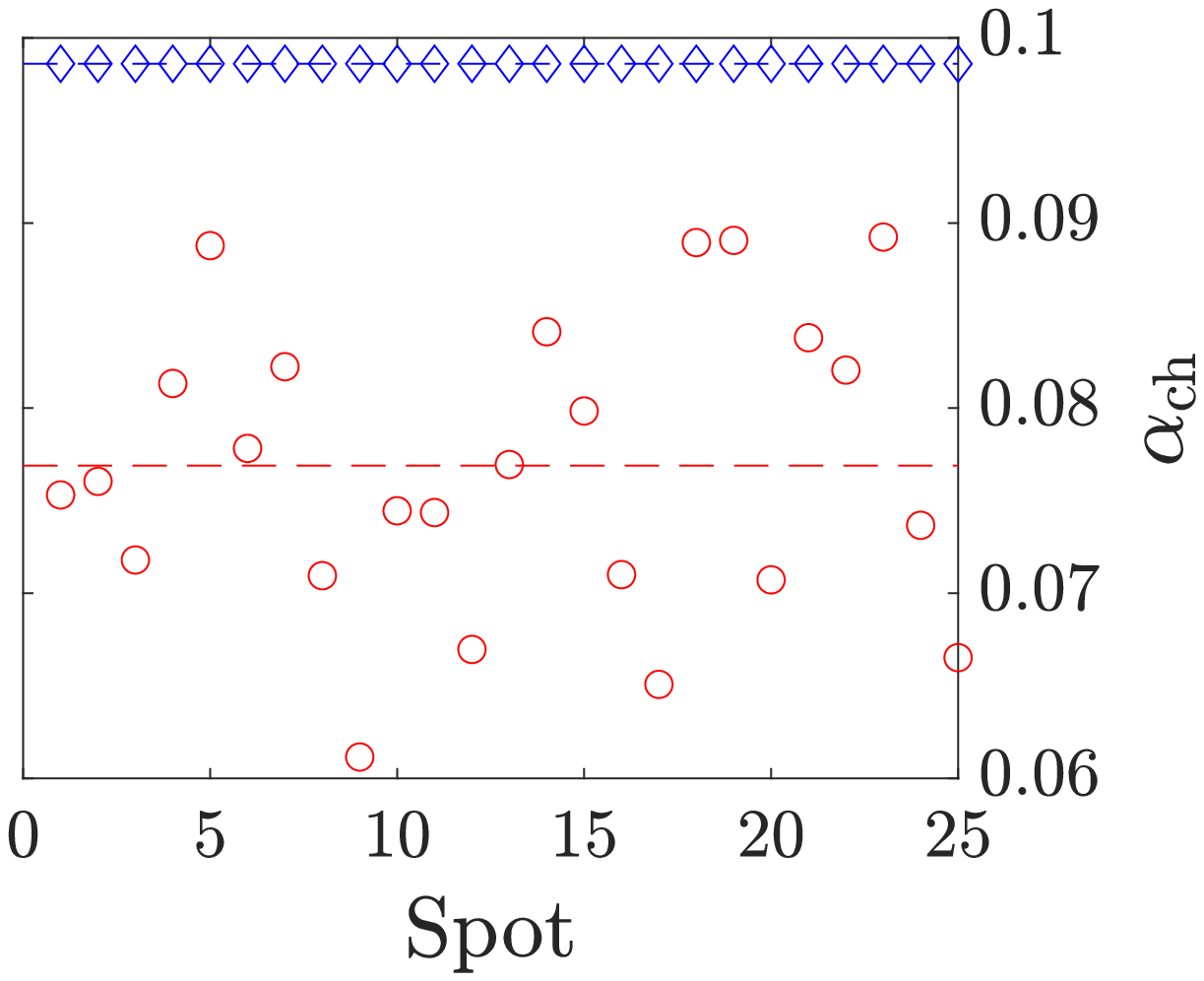}
	\caption{Comparison of identified absorption coefficients in case of 2-parameter (choroid coefficient free, red circles) and 1-parameter (choroid coefficient fixed, blue diamonds) estimation for eye number one. The dotted lines indicate the mean of the respective parameters over all spots.}
	\label{fig:fixed_vs_free}
\end{figure}

\noindent Next, we aim to analyze quantitatively and qualitatively the necessity of estimating two parameters, i.e., $\alpha_\text{RPE}$ and $\alpha_\text{ch}$. To this end, we compare the outcome of estimating both parameters simultaneously or fixing the choroid parameter to the empirical mean, i.e., $\alpha_\text{ch}=\bar{\alpha}_\text{ch}$ (see Table~\ref{t:mean_var}) and estimating only the $\alpha_\text{RPE}$ for eye number one. The choice of fixing $\alpha_\text{ch}$ and estimating only $\alpha_\text{RPE}$ is motivated by the findings in Sections~\ref{subsec:theosensi} and \ref{subsec:input_output_sensi}, where we found that the sensitivity with respect to perturbations of $\alpha_\text{RPE}$ is larger than with respect to perturbations of $\alpha_\text{ch}$. We show in Figure~\ref{fig:fixed_vs_free} the influence of fixing $\alpha_\text{ch}$ and only estimating of the remaining coefficient $\alpha_\text{RPE}$. As setting the choroid absorption $\alpha_\text{ch}$ to its mean value is an overestimation at this spot compared to estimating both parameters (see right of Figure~\ref{fig:fixed_vs_free}), the estimated absorption coefficient of the RPE $\alpha_\text{RPE}$ compensates this by being lower (left of Figure~\ref{fig:fixed_vs_free}).

\begin{figure}[!h]
	\centering
	\includegraphics[width=0.45\columnwidth]{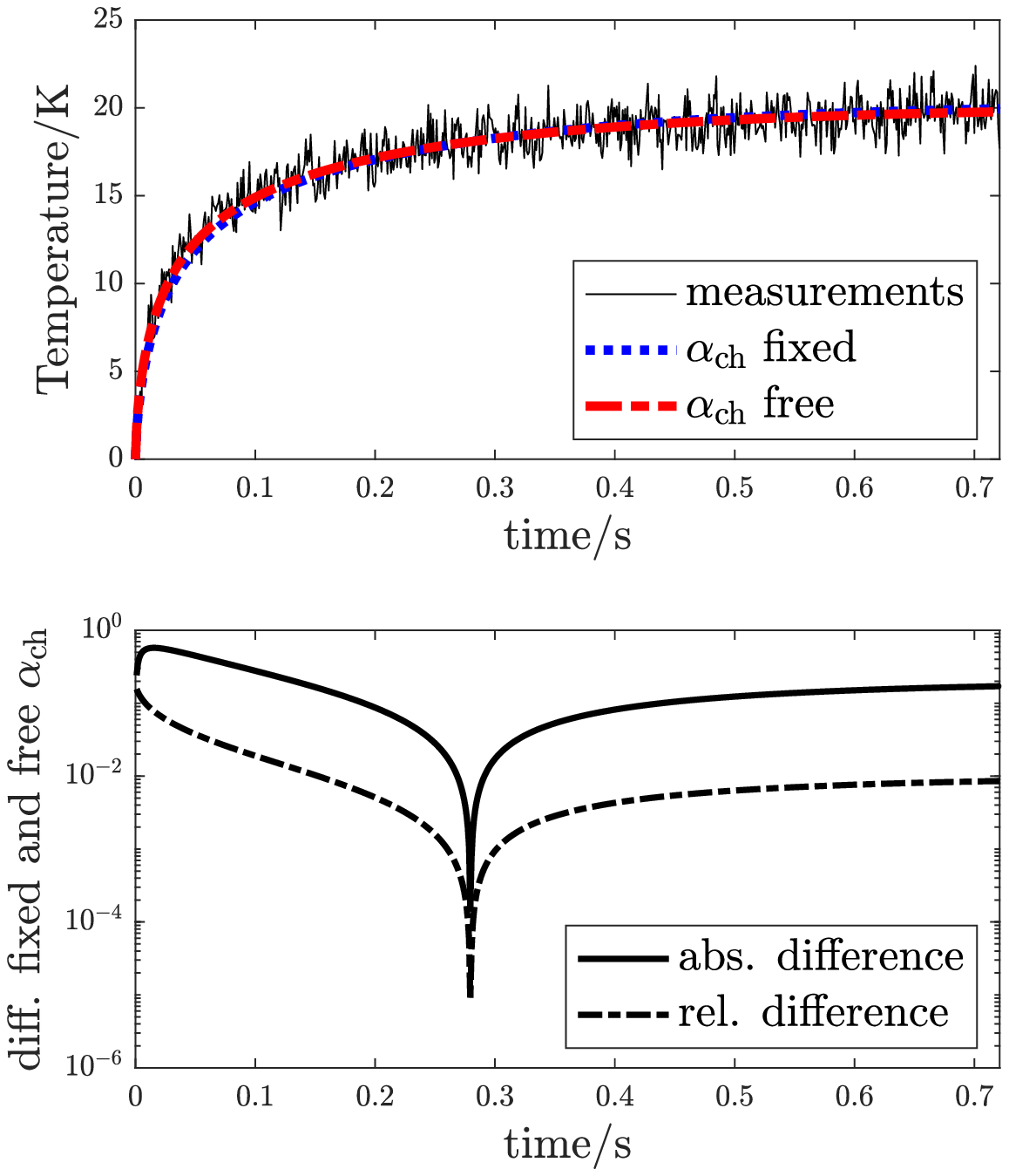}\hspace{-.7cm}
	\includegraphics[width=0.45\columnwidth]{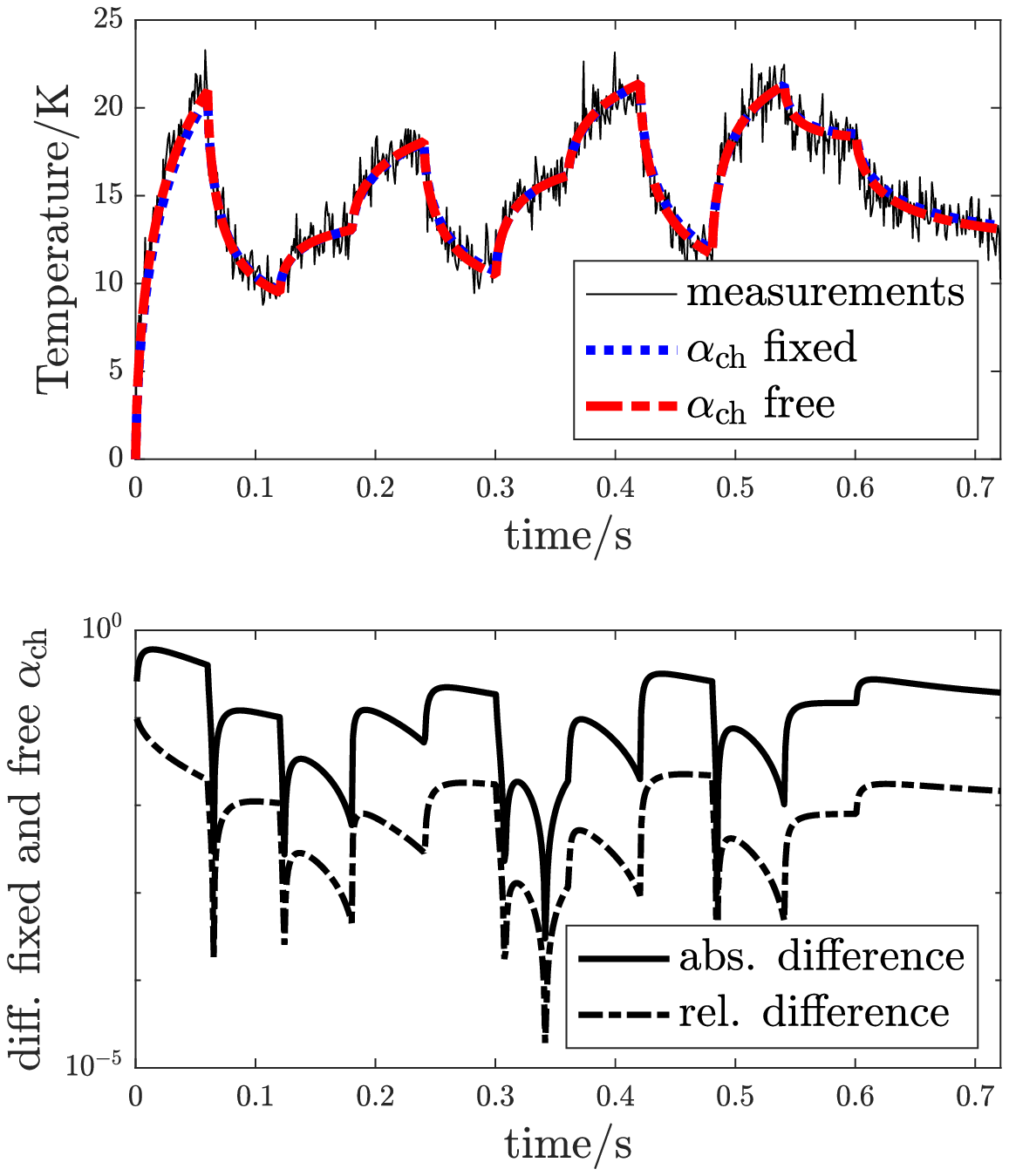}\\
	\caption{Comparison of volume temperatures for fixed $\alpha_\text{ch}$ and free $\alpha_\text{ch}$ for constant control (left) and the time-varying control \eqref{fig:fancyu} (right).}
	\label{fig:fixed_vs_free_outerr}
\end{figure}

	\noindent Having compared the effect of fixing one parameter to the empirical mean computed in the case study of Section~\ref{sec:case}, we now take a closer look at the corresponding outputs that would be result by those two different pairs of absorption coefficients, or in other words, we inspect the error in the output that is induced by only estimating one parameter. To this end, we compare in Figure~\ref{fig:fixed_vs_free_outerr} the outputs corresponding to the estimated absorption coefficients. We investigate the first spot of eye number one and compare two different controls, i.e. the constant laser power of 30\,mW (left), and the time-varying laser power depicted in Figure~\ref{fig:fancyu} (right). For either of the controls, we compared both volume and peak temperature, however we depict only the volume temperature, as both behave similarly. We see that, for both controls, the absolute error that is introduced in the volume temperature by only estimating $\alpha_\text{RPE}$ is below one. The same also holds for the peak temperature which is not shown here. In view of our application, the measurement noise, modeling, discretization and model reduction errors, we believe that this error can be acceptable. A detailed study of the effects of only identifying one parameter on the treatment outcomes in closed loop will be subject of future work.

\section{Parametric model reduction}
\label{sec:MOR}
\noindent
While the discretized PDE \eqref{eq:discretemodel} yields computational accuracy, the high state space dimension $n>80000$ is not suitable for parameter estimation and control in real-time. To this end, we investigate and compare two parametric model order reduction (MOR) approaches in our particular application in this section. More precisely, we first extend an approach based on Taylor expansion \cite{Kleyman2020b,Kleyman2021} and combine it with the interpolation-based parametric model order reduction method (pMOR) from \cite{Baur11}. Second, we consider a global basis (gb) approach \cite[Section 4.1]{benner2015survey} and pair it with a discrete empirical interpolation method (DEIM) \cite{Chaturantabut2010}.

%For instance, \cite{drohmann2011adaptive, amsallem2012nonlinear, eftang2012parameter, haasdonk2011training, peherstorfer2014localized, washabaugh2012nonlinear} are approaches to model order reduction related to our present context. 
We briefly provide some related works in the context of parametric model reduction for PDEs and refer the reader to the survey article \cite{benner2015survey} for further details. %In the works \cite{drohmann2011adaptive} and \cite{peherstorfer2014localized} a reduced basis approach for nonlinear PDEs is considered. %problem with some parameter. %\cite{drohmann2011adaptive} considered model order reduction of a system with discontinuous solution. 
In \cite{peherstorfer2014localized} a reduced basis approach with a localized discrete empirical interpolation method (LDEIM) is applied to compute several local subspaces, each adapted to a particular region of characteristic system behavior. In \cite{amsallem2012nonlinear}, instead of representing the solution in a fixed low dimensional subspace of global basis vectors, the authors present a MOR approach which approximates the solution in a low dimensional subspace generated by appropriately chosen local basis. The notion of hp empirical interpolation methods (EIM) was introduced in \cite{eftang2012parameter} to construct a partition of the parameter domain into parameter subdomains by means of so-called h-refinement. The EIM is applied independently on each subdomain to yield local approximation spaces by so-called p-refinement. In \cite{haasdonk2011training} and \cite{washabaugh2012nonlinear}, model order reduction using machine learning techniques was considered. In \cite{washabaugh2012nonlinear}, the authors constructed a model reduction framework based on the concept of local reduced-order basis, where in the offline phase, the local reduced-order bases were built using an unsupervised learning and in the online phase rank-one updates to the local bases were performed in order to increment accuracy.

\noindent The general framework of parametric model order reduction is as follows. Consider the dynamical control system with input space $\mathbb{R}^m$, output space $\mathbb{R}^l$ and state space $\mathbb{R}^n$ depending on a parameter $\alpha \in \mathcal{D}\subset \mathbb
R^q$,
\begin{align}
	\label{eq:dynamics}
		\dot x = Ax + B(\alpha)u, \qquad y = C(\alpha)x
\end{align}
with matrix $A\in\R^{n\times n}$, vector-valued functions $B: \mathcal{D} \rightarrow \R^{n\times m}$, $C: \mathcal{D} \rightarrow \R^{\ell\times n}$ and a control input $u\in\R^m$. Suppose that suitable projection matrices $V,W \in \R^{n\times d}$, $d\ll n$ with full rank are given. Then, we can define a  reduced-order model of dimension~$d$ by means of %$V\dot x_r = AVx_r + B(\alpha)u$, $y = C(\alpha)Vx_r$ or, equivalently, 
\begin{align}
\label{eq:reduced_system}
	\dot x_r = A_rx_r + B_r(\alpha)u \qquad y = C_r(\alpha)x_r,
\end{align}
where $A_r = W^\top AV$, $B_r(\alpha) = W^\top B(\alpha)$, and $C_r(\alpha) = C(\alpha)V$.

Since $B$ and $C$ depend nonlinearly on~$\alpha$, the reduced order surrogates $B_r(\alpha)$ and $C_r(\alpha)$ must be evaluated for each $\alpha$. The computational cost of such evaluation, however, depends on the original dimension $n$, as the high-dimensional nonlinearities $B(\alpha)\in \mathbb{R}^{n\times m}$ resp.\ $C(\alpha)\in \mathbb{R}^{l\times n}$ have to be evaluated and then projected by means of $V$ and $W$. Hence, whereas the first objective is the choice of suitable projection matrices $V$ and $W$, the second objective will be to address the latter issue by reducing complexity of the nonlinearity.

To this end, we compare two different parametric model reduction approaches: First, we pursue a Taylor series truncation of both $B(\alpha)$ and $C(\alpha)$ and pair it with parametric model-order reduction that was originally suggested for systems linear in the parameter in~\cite{Baur11} and subsequently extended to higher order polynomial approximations in our previous work~\cite{Kleyman2020b}. Second, we consider the well-established discrete empirical interpolation method (DEIM; \cite{Chaturantabut2010}) paired with a global basis approach \cite[Section~4.1.1]{benner2015survey}.

	For all following considerations, we will consider the stacked output operators of volume and peak temperature for the computation of the reduced order models, that is,
	\begin{align*}
	    C(\alpha) = \begin{pmatrix}C_{\text{vol}}(\alpha)\\ C_{\text{peak}}\end{pmatrix},
	\end{align*} as the former is important for estimation, whereas the latter is important for control, both of which have to be performed in real-time. Further, as parameter domain we choose \begin{align*}
	    \mathcal{D} & = [\bar\alpha_{\text{RPE}}-2\sigma_{\text{RPE}},\bar\alpha_{\text{RPE}}+2\sigma_{\text{RPE}}] \times[\bar\alpha_{\text{chor}}-2\sigma_{\text{chor}}, \bar\alpha_{\text{chor}}+2\sigma_{\text{chor}}] \\ & = [0.3821,1.1451]\times[0.0424,0.1549]
	\end{align*} 
	which represents a perturbation of the empirical mean by means of two empirical standard deviations in each direction as computed in the case study of Section~\ref{sec:case}, cf.\ Table~\ref{t:mean_var}.
\subsection{Taylor series truncation with pMOR}
\label{subsec:tay}
\noindent 
	In this subsection, we summarize the approach used in our previous work \cite{Kleyman2020a} to reduce the complexity of $B(\alpha)$ and $C_\text{vol}(\alpha)$ in~\eqref{eq:discreteevo} by a truncated Taylor expansion whereby obtaining vectors that depend polynomially on the absorption coefficients. We approximate the input operator ${B}(\alpha)$ and the output operators $\mathcal{C}_\text{vol}(\alpha)$ in \eqref{eq:discretemodel} by a Taylor series
\begin{align}
\label{def:taylor_expansion_b}
	B(\alpha) & \approx \sum_{i+j\leq \ktay}
	\left(\alpha_{\text{RPE}} - \alpha_{\text{RPE}}^0\right)^i \left(\alpha_{\text{chor}} - \alpha_{\text{chor}}^0\right)^jB_{i,j}(\alpha)\\
\label{def:taylor_expansion_c}
	C_{\text{vol}}(\alpha) & \approx \sum_{i+j\leq \ktay}
	\left(\alpha_{\text{RPE}} - \alpha_{\text{RPE}}^0\right)^i \left(\alpha_{\text{chor}} - \alpha_{\text{chor}}^0\right)^jC_{i,j}(\alpha),
\end{align}
where $\alpha\in\Dc$, $B_{i,j}\in\R^{n\times 1}$, $C_{i,j}\in\R^{2\times n}$, $i,j=0,\ldots,k$ and $\alpha^0=(\alpha_{\text{RPE}}^0, \alpha_{\text{chor}}^0)$ is the expansion point. 
The Taylor coefficients are
\begin{align*}
B_{i,j}(\alpha)  = \frac{1}{i!j!}\frac{\partial^{i+j} B(\alpha_0)}{\partial \alpha_{\text{RPE}}^i\alpha_{\text{chor}}^j}
\quad
\text{and}
\quad
C_{i,j}(\alpha)  = \frac{1}{i!j!}\frac{\partial^{i+j} C_{\text{vol}}(\alpha_0)}{\partial \alpha_{\text{RPE}}^i\alpha_{\text{chor}}^j}.
\end{align*}
This polynomial approximation is then paired with the pMOR approach \cite{Baur11,Kleyman2020b} using an Iterative Rational Krylov Algorithm (IRKA; \cite{gugercin2008h_2}) for the construction of $\mathcal{H}_2$-optimal projections $V$ and $W$. Due to the polynomial structure of the nonlinear parametric dependency, the evaluation of the nonlinearities only depends on the order of Taylor truncation order $k_\text{T}$ and the dimension of the reduced model $d$ and not on the full dimension $n$.

\subsection{Discrete empirical interpolation method with a global basis approach}
\label{subsec:deim}
\noindent
	An alternative approach used in the literature to obtain the projection matrices $V$ and $W$ is to sample the system at different parameter snapshots, to stack the resulting reduced bases into one matrix, and then to reduce it to a basis by means of singular value decomposition. This approach is called the global basis approach \cite[Section 4.1]{benner2015survey}.
	
	In order to efficiently evaluate the nonlinearities in this context, we use the discrete empirical interpolation method (DEIM) \cite{Chaturantabut2010}. The idea of DEIM is to approximate $B(\alpha)$ and $C(\alpha)$ by products of the form 
\begin{align}\label{eq:deim.1}
 B(\alpha)\approx 	\underbrace{U_B}_{n\times \kdeim}\underbrace{\widetilde{B}(\alpha)}_{\kdeim\times 1},\quad
 C(\alpha) \approx 	\underbrace{\widetilde C(\alpha)}_{2\times\kdeim}\underbrace{U_C}_{\kdeim\times n},
\end{align}
where $\kdeim \ll n$. Using these approximations, we have
\begin{align}
\label{eq:deim.vub}
	\underbrace{W^\top}_{d\times n}\underbrace{B(\alpha)}_{n\times 1} \approx \underbrace{W^\top U_B}_{d\times \kdeim}\underbrace{\widetilde B(\alpha)}_{\kdeim\times 1}\\
\label{eq:deim.cvu}
	\underbrace{C(\alpha)}_{1\times n}\underbrace{V}_{n\times d} \approx \underbrace{\widetilde C(\alpha)}_{1\times\kdeim}\underbrace{U_CV}_{\kdeim\times d},
\end{align}
where $U_C V$ and $W^\top U_B$ can be computed offline, and only the lower dimensional surrogates $\widetilde{B}(\alpha)\in \mathbb{R}^{\kdeim}$, resp.\ $\widetilde{C}(\alpha)\in \mathbb{R}^{\kdeim}$ have to evaluated online for a specific parameter $\alpha$.

	Following \cite{Chaturantabut2010}, we briefly describe how $U_B$ and $\widetilde{B}(\alpha)$ can be computed in order to approximate $B(\alpha) \approx U_B \widetilde{B}(\alpha)$ as in \eqref{eq:deim.1}. The reduction of $C(\alpha) \approx \widetilde{C}(\alpha)U_C$ can be computed completely analogously.
	
	First, we consider a discretization of the parameter domain $(\alpha_{1},\ldots,\alpha_{n_s})\subset\mathcal{D}$, $n_s\in \mathbb{N}$ and perform a singular value decomposition on the snapshot matrix, i.e., 
\begin{align}
	\label{eq:sing_value_b}
	\mathbf B 
	& := \left( B(\alpha_{1}),\ldots,B(\alpha_{n_s})\right)
	\in\R^{n\times n_s}\\\nonumber
	& = \hat V_B\hat S_B \hat W^\top_B
\end{align}
where $\hat V_B\in\R^{n\times r}$, $\hat W_B\in\R^{n_s\times r}$ and $\hat S_B = \diag(\sigma_1,\ldots,\sigma_r)$ with descending order and $r\leq\min\{n,n_s\}$ is the rank of $\mathbf B$. 

Then, $U_B$ is composed by the first $\kdeim$ columns of $\hat{V}_B$, where $\kdeim$ is a truncation parameter depending on the decay of the singular values. After this, we define the permutation matrix 
\begin{align}
\label{eq:defPB}
	P_B = \left[e_{j_1}, \ldots, e_{j_{\kdeim}}\right],\quad e_{j_i} \in \R^n 
\end{align}
whose columns $e_{j_i}$ are a permutation of $k_D$ elements of the standard basis. A suitable choice of indices is performed by means of an adaptive algorithm \cite[Algorithm 1]{Chaturantabut2010}.

The approximation $\widetilde{B}(\alpha)$ can now be obtained by solving a projected version of \eqref{eq:deim.1}, i.e.,
\begin{align}
	\label{eq:deim.permutation}
	\widetilde B(\alpha) = \left(P_B^\top U_B\right)^{-1}P_B^\top B(\alpha).
\end{align}
Thus, the indices in $P_B$ select the interpolation points consisting of particular rows of $B(\alpha)$ that are then combined by means of $U_B$.
Analogously, we compute the corresponding counterparts to obtain $\widetilde{C}(\alpha)$ by means of a snapshot matrix \linebreak${\mathbf{C} = \left( C(\alpha_{1}),\ldots,C(\alpha_{n_s})\right)}$, a selection of right singular vectors $U_C$ and a permutation matrix $P_C$.

	Since $P_B^\top$ only chooses and permutes $\kdeim$ rows of $B(\alpha)$ in \eqref{eq:deim.permutation}, the assembly of $\widetilde B(\alpha) = P_B^\top B(\alpha)$ may be implemented efficiently and independently of the original dimension. The same also holds true for the output operator $\widetilde{C}(\alpha) = C(\alpha)P_C$.

	In the following, we investigate and compare the presented DEIM- and Taylor-based approaches of this and the previous Subsection~\ref{subsec:tay}. To this end, we will consider various reduction and truncation orders $d,\kdeim$ and $\ktay$ that represent the design parameters of the MOR methods. Moreover, we will consider separately the case of one parameter $\alpha_\text{RPE}$, i.e., we fix the value of $\alpha_{\text{ch}}$ in Subsection~\ref{sec:MOR_fixed} and the case of two independent parameters $(\alpha_{\text{ch}},\alpha_{\text{RPE}})$ in Subsection~\ref{sec:MOR_two}.

%%%%%%%%%%%%%%%%%%%%%%%%%%%%%%%%%%%%%%%%%%%%%%%%%%%%%%%%%%%

\subsection{Comparison in case the choroid absorption is fixed: one parameter}	
\label{sec:MOR_fixed}
\noindent
	We concluded in Section~\ref{sec:sensi} that the sensitivity w.r.t.\ the absorption coefficient in the choroid is lower than w.r.t.\ its counterpart in the RPE and that fixing this parameter leads to relatively small errors, cf.\ Figure~\ref{fig:fixed_vs_free_outerr}. Therefore, we first investigate the two presented MOR-techniques in case of fixing the absorption coefficient in the choroid to its empirical mean as obtained in the case study of Section~\ref{sec:case}, that is $\alpha_\text{ch}=\bar{\alpha}_\text{ch}$ as given in Table~\ref{t:mean_var}. 

\begin{figure}[h]
	\centering
	\includegraphics[width=.6\columnwidth]{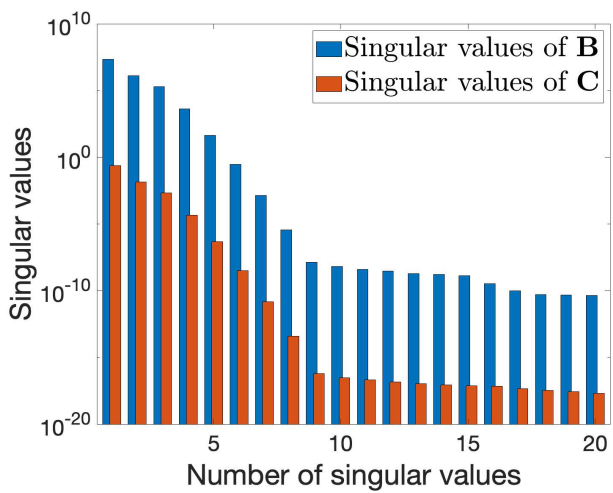}
    \caption{First 20 singular values of $\mathbf B$ and $\mathbf C$ for DEIM with one parameter, cf.\ \eqref{eq:sing_value_b} using $n_s=20$ snapshots.}
	\label{fig:deim_sing_values_1p}
\end{figure}

First, we inspect in Figure~\ref{fig:deim_sing_values_1p} the singular values of $\mathbf{B}$ in \eqref{eq:sing_value_b} and the corresponding counterpart for the output $\mathbf{C}$. In both cases, the decay in the singular values flattens after the first eight values. However, truncating the singular value decomposition after the third value results in and the relative cumulative energy content $\sum_{i=1}^{3}\sigma_i^2/\sum_{i=1}^{n}\sigma_i^2= 0.9998$ for both $\mathbf B$ and $\mathbf C$.
	
\noindent
We will now compare the resulting errors for trajectories of the reduced system over the time and parameter domain.
To ensure comparability of the errors at different absorption parameters, we always apply the constant control corresponding to the steady-state output of 30\,K, i.e., for a given value of the parameter $\alpha$, we choose the constant control
\begin{align}
\label{eq:sscontrol}
    u \equiv -30 (C_{\text{peak}} A^{-1}B(\alpha) )^{-1}.
\end{align}
As initial value, we consider $x(0)=0$ as due to linearity only temperature increases w.r.t.\ the ambient temperature are modeled.

Denoting by $y_*(\alpha,\cdot) = C_{r,*}(\alpha)x_r(\alpha,\cdot)$ the output trajectory of the reduced order model (obtained from applying model reduction as described in Subsections~\ref{subsec:tay} or \ref{subsec:deim}) and by $y_{\text{f},*}(\alpha,\cdot) = C_*(\alpha)x(\alpha,\cdot)$ the output trajectory of the full-order model \eqref{eq:discretemodel}, where $*\in\{\text{vol},\text{peak}\}$, we will compare the MOR error by means of two different measures:
\begin{align}
\label{eq:maxerr}
    \text{err}_{\infty} & = \max_{\alpha,i}\left\{\frac{\vert y_*(\alpha,i) - y_{\text{f},*}(\alpha,i)\vert}{\vert y_{\text{f},*}(\alpha,i)\vert} \right\},\\
     \label{eq:L2err}
    \text{err}_{\infty,2} & = \max_{\alpha}\left\{\left(\frac{\sum_i\vert y_*(\alpha,i) - y_{\text{f},*}(\alpha,i)\vert^2}{\sum_{i}\vert y_{\text{f},*}(\alpha,i)\vert^2}\right)^{1/2} \right\}
\end{align}
In the parameter domain, we always consider the worst case, i.e., the maximal error, as an over- or undertreatment at one single spot is already undesirable in terms of our application. In time, however, we consider both the maximal and the $L_2$ error, as a deviation might be less critical if it happens for a short amount of time.

In the following tables, the symbol $\dagger$ will denote an unsuccessful model reduction due to, e.g., numerical instabilities that occur in the computations, such as a failed determination of a Cholesky factor required for pMOR in the Taylor-based approach \cite{Kleyman2020b} or an unstable reduced model.

\begin{table}[h]
    \begin{center}
            \begin{tabular}{ | c || c | c | c | c | }
            	\hline
            	\multicolumn{5}{|c|}{Volume temperature} \\
            	\hline
            	$\ktay\setminus d$ & 5 & 6 & 7 & 8\\
            	\hline\hline
            	2 & 0.2495 & 0.2467 & 0.2400 & $\hspace{0.41cm}\dagger\hspace{0.41cm}$ \\
            	\hline
            	3--10 & 0.2532 & 0.2505 & 0.2437 & $\dagger$ \\
            	\hline
        \end{tabular}
                \begin{tabular}{ | c || c | c | c | c |}
            	\hline
            	\multicolumn{5}{|c|}{Peak temperature} \\
            	\hline
            	$\ktay\setminus d$ & 5 & 6 & 7 & 8\\
            	\hline\hline
            	2 & 0.2235 & 0.2207 & 0.2230 & $\hspace{0.41cm}\dagger\hspace{0.41cm}$ \\
            	\hline
            	3 & 0.2262 & 0.2235 & 0.2258 & $\dagger$ \\
            	\hline
            	4--10 & 0.2260 & 0.2233 & 0.2256 & $\dagger$ \\
            	\hline
        \end{tabular}\\
        \vspace{.2cm}
        \begin{tabular}{ | c || c | c | c | c |}
            	\hline
            	\multicolumn{5}{|c|}{
            	Volume temperature} 
              	\\
            	\hline
            	$\kdeim\setminus d $& 5 & 6 &7&8\\
            	\hline\hline
            	3 & 0.0331 & 0.0243 & 0.0184 & 0.0174\\
            	\hline
            	4--10 & 0.0336  & 0.0248 & 0.0189 & 0.0179 \\
            	\hline
            \end{tabular}
            \begin{tabular}{ | c || c | c | c | c |}
            	\hline
            	\multicolumn{5}{|c|}{
            	Peak temperature} 
              	\\
            	\hline
            	$\kdeim\setminus d $& 5 & 6 &7&8\\
            	\hline\hline
            	3 & 0.0357 &  0.0079 & 0.0069 & 0.0057\\
            	\hline
            	4--10 & 0.0355  & 0.0077 & 0.0067 & 0.0055 \\
            	\hline
            \end{tabular}
            \caption{Comparison of maximal relative error $\text{err}_{\infty}$, cf.\ \eqref{eq:maxerr} in volume and peak temperature in the one-parameter case.}
	\label{table:max_relative_error}
        \end{center}
\end{table}

%% inspect max error
\noindent In Table~\ref{table:max_relative_error}, we show the maximum of the relative error, cf.\ \eqref{eq:maxerr}, in the volume and the peak temperature for various combinations of both $\kdeim$ resp.\ $\ktay$ with varying projection orders $d$. 
 We see that the DEIM-based approach clearly performs better over all considered orders by one order of magnitude. %Moreover, we observe that the error of the DEIM-based appraoch is monotonically decreasing for increased projection order $d$ and DEIM order $\kdeim$.

\begin{table}[h]
    \begin{center}
    \begin{tabular}{ | c || c | c | c | c |}
        \hline
        \multicolumn{5}{|c|}{Volume temperature} \\
        \hline
        $\ktay\setminus d$ & 5 & 6 & 7 & 8 \\
        \hline\hline
        2 & 0.5841 & 0.5531 & 0.5409 & $\dagger$\\
        \hline
        3 & 0.4472 & 0.4623 & 0.4298 & $\hspace{0.41cm}\dagger\hspace{0.41cm}$ \\
        \hline
        4 & 0.3726 & 0.3879 & 0.3633 & $\dagger$ \\
        \hline
        5 & 0.3275 & 0.3472 & 0.3218 & $\dagger$ \\
        \hline
        6 & 0.2953 & 0.2891 & 0.2916 & $\dagger$ \\
        \hline
        7 & 0.2711 & 0.2660 & 0.2687 & $\dagger$ \\
        \hline
        8 & 0.2521 & 0.2476 & 0.2504 & $\dagger$ \\
        \hline
        9 & 0.2365 & 0.2326 & 0.2355 & $\dagger$ \\
        \hline
    \end{tabular}
    \begin{tabular}{ | c || c | c | c | c |}
        \hline
        \multicolumn{5}{|c|}{Peak temperature} \\
        \hline
        $\ktay\setminus d$ & 5 & 6 & 7 & 8 \\
        \hline\hline
        2 & 0.3079 & 0.2970 & 0.2817 & $\dagger$ \\
        \hline
        3 & 0.2351 & 0.2155 & 0.2219& \hspace{0.41cm}$\dagger$\hspace{0.41cm} \\
        \hline
        4 & 0.1965 & 0.1871 & 0.1882 & $\dagger$ \\
        \hline
        5 & 0.1727 & 0.1582 & 0.1667 & $\dagger$ \\
        \hline
        6 & 0.1557 & 0.1549 & 0.1511 & $\dagger$ \\
        \hline
        7 & 0.1430 & 0.1425 & 0.1392 & $\dagger$ \\
        \hline
        8 & 0.1329 & 0.1326 & 0.1298 & $\dagger$ \\
        \hline
        9 & 0.1247 & 0.1246 & 0.1220 & $\dagger$ \\
        \hline
    \end{tabular}\\
    \vspace{.2cm}
    \begin{tabular}{| c || c | c | c | c |}
        \hline
        \multicolumn{5}{|c|}{Volume temperature} \\
        \hline
        $\kdeim\setminus d$ & 5 & 6 & 7 & 8 \\
        \hline\hline
        3 & 0.0029 & 0.0023 & 0.0019 & 0.0019 \\
        \hline
        4 & 0.0025 & 0.0020 & 0.0017 & 0.0017\\
        \hline
        5 & 0.0022 & 0.0018 & 0.0015 & 0.0015 \\
        \hline
        6 & 0.0019 & 0.0016 & 0.0013 & 0.0013 \\
        \hline
        7 & 0.0018 & 0.0015 & 0.0012 & 0.0012 \\
        \hline
        8 & 0.0017 & 0.0014 & 0.0011 & 0.0011 \\
        \hline
        9 & 0.0016 & 0.0013 & 0.0011 & 0.0011 \\
        \hline
%        10 & 0.0015 & 0.0012 & 0.0010 & 0.0010 & 0.0009\\
    \end{tabular}
    \begin{tabular}{| c || c | c | c | c |}
        \hline
        \multicolumn{5}{|c|}{Peak temperature} \\
        \hline
        $\kdeim\setminus d$ & 5 & 6 & 7 & 8 \\
        \hline\hline
        3 & 0.0015 & 0.0010 & 0.0008 & 0.0008 \\
        \hline
        4 & 0.0012 & 0.0008 & 0.0007 & 0.0006 \\
        \hline
        5 & 0.0010 & 0.0007 & 0.0006 & 0.0006 \\
        \hline
        6 & 0.0009 & 0.0006 & 0.0005 & 0.0005 \\
        \hline
        7 & 0.0009 & 0.0006 & 0.0005 & 0.0005 \\
        \hline
        8 & 0.0008 & 0.0005 & 0.0005 & 0.0004 \\
        \hline
        9 & 0.0008 & 0.0005 & 0.0004 & 0.0004 \\
        \hline
%        10 & 0.0007 & 0.0005 & 0.0004 & 0.0004 & 0.0004\\
    \end{tabular}
    \caption{Comparison of relative $L^2$-error $\text{err}_{\infty,2}$, cf.\ \eqref{eq:L2err}, in volume temperature and peak temperature for the one-parameter case.}
	\label{table:relative_L2}
	\end{center}
\end{table}
%% inspect L2 error
\noindent In Table~\ref{table:relative_L2}, we compare the relative $L_2$-error computed via~\eqref{eq:L2err}. Here, the DEIM-based approach clearly outperforms the Taylor-based approach by approximately two orders of magnitude for both the peak and the volume temperature. Further, we see that both approaches have decreasing $L_2$ errors for increased MOR order $d$ and DEIM order $\kdeim$ resp. Taylor truncation order $\ktay$. 

%Again, we first marked the best values of the Taylor-based approach in gray and then correspondingly the best values of DEIM with the same order.
%We observe that for the Taylor-based appraoch the errors in the volume temperature are relatively constant along the Taylor degree $\ktay$ and the MOR order $d$, whereas, for the peak temperature, only the former is true and the error decreases for increased MOR order.
%On the other hand, we observe that the corresponding error of the DEIM-based approach, is smaller by two orders of magnitude along all choices of $\kdeim$ and $d$. Again, the error is monotonically decreasing in both $\kdeim$ and $d$.
%% our conclusion
Thus, we conclude that both, in terms of volume and peak temperature and for both performance measures \eqref{eq:L2err} and \eqref{eq:maxerr}, the DEIM-based approach of Subsection~\ref{subsec:deim} is better suited for our application than the Taylor-based approach of Subsection~\ref{subsec:tay} that was used in previous works \cite{Kleyman2020b,Kleyman2021}.

The reason for the poor performance of the Taylor-based approach in both error measures \eqref{eq:L2err} and \eqref{eq:maxerr} is due to the nature of the Taylor approximation: We obtain a relatively good approximation at the expansion points, however the errors towards the boundaries of the parameter domains become very large, leading to a large maximal error over the parameter domain. %Thus, the maximal error over the parameter domain is always attained at the bounds.
%%%%%%%%%%%%%%%%%%%%%%%%%%%%
\subsection{Comparison in case the choroid absorption is free: two parameters}
\label{sec:MOR_two}
\noindent 
Here, we will investigate the case of two independent parameters $\alpha = (\alpha_\text{RPE},\alpha_\text{ch}$).
To this end, we will proceed analogously to the one-parameter case in the previous Subsection~\ref{sec:MOR_fixed} and again use the constant steady-state control as input, cf.\ \eqref{eq:sscontrol} as well as the comparison metrics $\text{err}_{\infty}$ and $\text{err}_{\infty,2}$ as defined in \eqref{eq:maxerr} and \eqref{eq:L2err}, respectively.

\begin{figure}[h]
	\centering
	\includegraphics[width=.6\columnwidth]{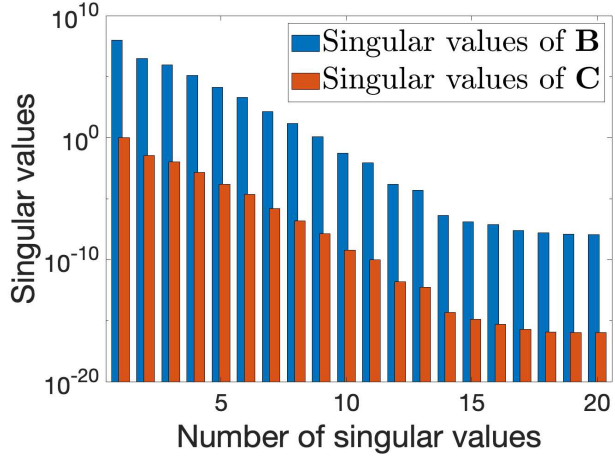}
    \caption{First 20 singular values of $\mathbf B$ and $\mathbf C$ for DEIM with two parameters, cf.\ \eqref{eq:sing_value_b} using $n_s=20$ snapshots.}
	\label{fig:deim_sing_values_2p}
\end{figure}

	In Fig. \ref{fig:deim_sing_values_2p}, the singular values of the snapshot matrices $\mathbf B$ and $\mathbf C$ used in the DEIM-based approach are shown. In both cases, the decay in the singular values is exponential. Already after four singular values, the cumulative relative energy content reaches $\sum_{i=1}^{4}\sigma_i^2/\sum_{i=1}^{n}\sigma_i^2 = 0.9998$ for both $\mathbf B$ and $\mathbf C$.
 \noindent In Table~\ref{t:max2} we depict the maximal error $\text{err}_\infty$, cf.\ \eqref{eq:maxerr}, in peak and volume temperature for varying orders $d$ and $\kdeim$ resp.\ $\ktay$. Similar to the one-parameter case, the DEIM-based approach leads to smaller errors, here with more than one order of magnitude.
 
 Considering the $L_2$ error $\text{err}_{\infty,2}$, cf.\ \eqref{eq:L2err}, in Table~\ref{t:L2two}, the DEIM-based approach achieves an error that is even smaller by approximately two orders of magnitude.

Thus, we conclude that also in the two-parameter case, the DEIM-based approach of Subsection~\ref{subsec:deim} is the method of choice in view of our application.

	%In Table~\ref{t:L2two}, we compare the relative $L^2$-error. Since the lowest relative $L^2$-error in Taylor truncation happens for projection order 6 and there is no significant difference along the projection order of pMOR, we recommend the projection order 6 for the pMOR, which is a choice also supported by the maximum relative error. As for DEIM, there is no drastic change in error for $\kdeim \geq 3$ for a fixed projection order. For this reason, $\kdeim = 3$ would be sufficient. In this row ($\kdeim = 3$), the projection order 7 gives the second lowest relative $L^2$-error with $0.0012$, which is decreasing only gradually. 

\begin{table}[h]
	\centering
        \begin{tabular}{ | c || c | c | c | c |}
		\hline
		\multicolumn{5}{|c|}{Volume temperature} \\
	        	\hline
		$\ktay\setminus d$ & 5 & 6 & 7 & 8\\
		\hline\hline
        3  & 0.4315 & 0.4357 &  0.4321 & 0.4249 \\
        \hline
        4  & 0.4259 & 0.4326 & 0.4286 & 0.4283 \\
        \hline
        5  & 0.4326 & 0.4359 & 0.4308 & 0.4306 \\
        \hline
        6  & 0.4262 & 0.4327 & 0.4290 & 0.4288 \\
        \hline
		7--10& $\dagger$ & $\dagger$ & $\dagger$ & $\dagger$ \\
        \hline
	\end{tabular}
	\begin{tabular}{ | c || c | c | c | c |}
		\hline
		\multicolumn{5}{|c|}{Peak temperature} \\
		\hline
		$\ktay\setminus d$ & 5 & 6 & 7 & 8 \\
		\hline\hline
		3 & 0.3425 & 0.3409 &  0.3412 & 0.3442 \\
		\hline
		4 & 0.3455 & 0.3439 & 0.3442 & 0.3423 \\
		\hline
		5 & 0.3475 & 0.3453 & 0.3442 & 0.3425 \\
		\hline
		6 & 0.3449 & 0.3428 & 0.3430 & 0.3419 \\
		\hline
		7--10& $\dagger$ & $\dagger$ & $\dagger$ & $\dagger$ \\
		\hline
	\end{tabular}\\
	\vspace{0.2cm}
		\begin{tabular}{ | c || c | c | c | c |}
		\hline
		\multicolumn{5}{|c|}{Volume temperature} \\
	        	\hline
		$\kdeim\setminus d$ & 5 & 6 & 7 & 8 \\
		\hline\hline
        3 & 0.0171 & 0.0161 & 0.0156 &  0.0159 \\
        \hline
        4 & 0.0144 & 0.0120 & 0.0100 & 0.0103 \\
        \hline
        5 & 0.0166 & 0.0139 & 0.0116 & 0.0117 \\
        \hline
        6--10 & 0.0166 & 0.0140 & 0.0116 & 0.0118 \\
        \hline
	\end{tabular}
		\begin{tabular}{ | c || c | c | c | c |}
		\hline
		\multicolumn{5}{|c|}{Peak temperature} \\
	        	\hline
		$\kdeim\setminus d$ & 5 & 6 & 7 & 8 \\
		\hline\hline
		3 & 0.0137 &  0.0061 & 0.0052 & 0.0045 \\
		\hline
		4 & 0.0123 & 0.0081 & 0.0061 & 0.0062 \\
		\hline
		5 & 0.0130 & 0.0060 & 0.0050 & 0.0042 \\
		\hline
		6--10 & 0.0129 & 0.0060 & 0.0050 & 0.0041 \\
		\hline
	\end{tabular}
	\caption{Comparison of maximal relative error $\text{err}_{\infty}$, cf.\ \eqref{eq:maxerr} in volume and peak temperature in the two-parameter case.}
	\label{t:max2}
\end{table}

\begin{table}[h]
	\centering
	\begin{tabular}{|c||c|c|c|c|}
			\hline
			\multicolumn{5}{|c|}{Volume temperature} \\
			\hline
			$\ktay \setminus d$ & 5 & 6 & 7 & 8 \\
			\hline\hline
            3 & 0.0107 & 0.0104 & 0.0101 & 0.0100 \\
			\hline
            4 & 0.0089 & 0.0087 & 0.0086 & 0.0084 \\
			\hline
            5 & 0.0079 & 0.0077 & 0.0076 & 0.0075 \\
			\hline
            6 & 0.0071 & 0.0070 & 0.0069 &  0.0068 \\
			\hline
			7--10 & $\dagger$ & $\dagger$ & $\dagger$ & $\dagger$ \\
			\hline
		\end{tabular}
	\begin{tabular}{|c||c|c|c|c|}
			\hline
			\multicolumn{5}{|c|}{Peak temperature} \\
			\hline
			$\ktay \setminus d$ & 5 & 6 & 7 & 8\\
			\hline\hline
			3 & 0.0057 & 0.0055 & 0.0054 & 0.0053 \\
			\hline
			4 & 0.0048 & 0.0047 & 0.0046 & 0.0046 \\
			\hline
			5 & 0.0042 & 0.0042 & 0.0041 & 0.0040 \\
			\hline
			6 & 0.0038 & 0.0038 & 0.0037 &  0.0037 \\
			\hline
			7--10 & $\dagger$ & $\dagger$ & $\dagger$ & $\dagger$ \\
			\hline
		\end{tabular}\\
		\vspace{0.2cm}
		\begin{tabular}{|c||c|c|c|c|}
			\hline
			\multicolumn{5}{|c|}{Volume temperature ($\times1$e$^{-3}$)} \\
			\hline
			$\kdeim \setminus d$ & 5 & 6 & 7 & 8 \\
			\hline
			3 & 0.1165 &  0.0956 & 0.0779 & 0.0760 \\
			\hline
			4 & 0.0876 & 0.0731 & 0.0600 & 0.0593 \\
			\hline
			5 & 0.0738 & 0.0607 & 0.0483 & 0.0478 \\
			\hline
			6 & 0.0666 & 0.0549 & 0.0438 & 0.0434 \\
			\hline
			7 & 0.0611 & 0.0505 & 0.0403 & 0.0401 \\
			\hline
			8 & 0.0568 & 0.0470 & 0.0376 & 0.0374 \\
			\hline
			9 & 0.0533 & 0.0442 & 0.0353 & 0.0352 \\
			\hline
			10 & 0.0504 & 0.0418 & 0.0335 & 0.0333 \\
			\hline
			\end{tabular}
			\begin{tabular}{|c||c|c|c|c|}
			\hline
			\multicolumn{5}{|c|}{Peak temperature ($\times1$e$^{-4}$)} \\
			\hline
			$\kdeim \setminus d$ & 5 & 6 & 7 & 8 \\
			\hline
			3 & 0.2561 &  0.1511 & 0.1203 & 0.1125 \\
			\hline
			4 & 0.2153 & 0.1313 & 0.1071 & 0.1026 \\
			\hline
			5 & 0.1854 & 0.1102 & 0.0880 & 0.0841 \\
			\hline
			6 & 0.1672 & 0.0997 & 0.0797 & 0.0764 \\
			\hline
			7 & 0.1535 & 0.0917 & 0.0735 & 0.0705 \\
			\hline
			8 & 0.1427 & 0.0853 & 0.0685 & 0.0658 \\
			\hline
			9 & 0.1339 & 0.0802 & 0.0644 & 0.0619 \\
			\hline
			10 & 0.1266 & 0.0758 & 0.0610 & 0.0587 \\
			\hline
			\end{tabular}
	\caption{Comparison of relative $L^2$-error $\text{err}_{\infty,2}$, cf.\ \eqref{eq:L2err}, in volume temperature and peak temperature for the two-parameter case.}
\label{t:L2two}
\end{table}

%%%%%%%%%%%%%%%%%%%%%%%%%%%%%%

%\subsection{Conclusion of the model comparison}
%We conclude the in case of one parameter, the DEIM-based approach of Subsection~\ref{subsec:deim} clearly outperforms the Taylor-based approach of Subsection~\ref{subsec:tay} that was suggested and used in previous work~\cite{Kleyman2020b,Kleyman2021}.

%\noindent We note that in view of our application, a small projection order $d$ is more important than a small Taylor truncation order $\ktay$ or DEIM order $\kdeim$. This is because the latter two only enter the evaluation of the input operator $\tilde{B}(\alpha)$ and the volume temperature output $\tilde{C}(\alpha)$, whereas the former governs the dimension of the optimal control problem, that we aim to use for model predictive control, cf.\ Section~\ref{sec:mpc}.

%\noindent
%1p: Taylor: 6,6; Deim: 6,3
%2p: Taylor 7,3; Deim: 6,3

%Hence, we choose the DEIM-based approach with $\kdeim=3$ and projection order $d=6$ (boxed values in the respective tables) as this choice induces an error that is suitable for our application, e.g., in terms of the maximal error metric $\text{err}_{\infty}$ a <2.5\% maximal relative error in volume temperature, and a $1$\% maximal error in peak temperature. The values in the $L_2$-metric $\text{err}_{\infty,2}$ are even better, cf.\ Table~\ref{table:relative_L2}.

\section{Real-time capability of MPC using the reduced order model}
\label{sec:mpc}
\noindent Here, we briefly provide resulting computation times of solving the optimal control problem (OCP) that has to be solved to compute an MPC-feedback. We use a reduced model obtained from the DEIM-based approach with dimension $d=6$ and order $\kdeim =3$. Consider the absorption coefficient prefactor $\alpha$, an initial date $x^0$, a reference peak temperature $y_\text{peak,ref}=30$\,K (effectivity of the treatment), a maximal peak temperature $y_\text{peak,max}=32$\,K (safety of the treatment), the steady state control $u_\text{ref}$ computed by means of \eqref{eq:sscontrol} and a maximal laser power $u_\text{max}=0.1$\,W. For a prediction horizon $N\in \mathbb{N}$, $N \geq 2$, we consider the optimal control problem (OCP)
\begin{alignat}{2}
\nonumber
                \min_{u\in \mathbb{R}^{N}} \sum_{k=0}^{N-1}  &|C_{\text{peak},r}x_k -y_\text{peak,ref}|^2 + &&5\cdot 10^4|u_k-u_{\text{ref}}|^2 \\
\nonumber
                \text{s.t. } x_{k+1} &= A_r x_k + B_r(\alpha)u_k  && k=0,\ldots,N-1\\
\label{eq:OCP}
                x_0 &= x^0 \\
\nonumber
                0&\leq u_k \leq u_\text{max} &&k=0,\ldots, N-1\\
\nonumber
                C_{\text{peak},r}x_k &\leq y_\text{peak,max} && k=0,\ldots, N-1.
\end{alignat}
We show in Table~\ref{t:comptimes} the computation times needed to solve the above OCP in an MPC-controller for different prediction horizons $N$ with a closed-loop length of 20 sampling instances. The computations were performed on a MacBook Pro with a 6-Core Intel Core i7\,@\,2.6\,GHz and 32\,GB RAM by means of a C++-implementation using the OCP-solver OSQP \cite{osqp}. Starting from the second MPC iteration, we utilize a suitable warm-start as common in MPC, cf.\ \cite[Section 10.5]{Gruene2016b} using the optimal solution of the previous MPC iteration as an initial guess for the OCP-solver. The computation times shown in Table~\ref{t:comptimes} show that the low-dimensional surrogate model allows for a fast online implementation of an MPC algorithm with a repetition rate of 1\,kHz. This would not be possible for the full model with state dimension 80 000, as solving \eqref{eq:OCP} using a model with state dimension 800 already requires approx.\ 82\,ms. Last, we mention that the maximal computation time is always achieved in the first MPC iteration, where currently no warm start is used. Appropriate choices, such as the state emanating from the steady-state control $u_\text{ref}$, will be considered in the future.
\begin{table}[H]
	\centering
	\begin{tabular}{|c||c|c|c|c|c|}
	\hline
	$N$& $2$ & $5$ & $10$ & $15$ & $20$\\\hline\hline
		avg.\ time\,(ms)&0.06 &0.16 &0.29& 0.37&0.41\\
		max.\ time\,(ms)&0.12 &0.34 &0.59& 0.65&0.67\\
		\hline
	\end{tabular}
		\caption{Average and maximal computation time for solving \eqref{eq:OCP} in an MPC-controller.}
	\label{t:comptimes}
\end{table}

\section{Conclusion}
\noindent	We carried out parameter estimation and computed parametric model order reduction for real-time model-based control in retinal laser treatment. In the first part of the paper, a case study of the absorption coefficients in porcine eyes was conducted. In the case study, the parameter range of the absorption coefficients was estimated by an optimization-based identification method on the model described by the heat equation, which depends nonlinearly on two unknown absorption coefficients. Furthermore, we identified the dominant parameter through a qualitative and quantitative sensitivity analysis in time and frequency domain, and compared the resulting output error when only identifying one absorption parameter. 
	
	In the subsequent part of the paper, we compared two state-of-the art parametric model reduction schemes with the goal of model predictive control in real-time based on the empirical range of the parameters. Models were designed based on two techniques for a fixed value of the absorption coefficient in the choroid and the two absorption coefficients as independent parameters and various orders of the models were compared in simulated output error.
	We found that the discrete empirical interpolation approach paired with a global basis outperforms a recently proposed model reduction based on Taylor approximation, in both cases of estimating one and two parameters. Lastly, we showcased that the obtained low-dimensional model enables us to perform MPC with very high sampling rates of 1\,kHz.

\bibliographystyle{abbrv} 
\bibliography{references}

\appendix
\section{Some remarks on the confidence intervals}
\label{sec:appendix}
\noindent We briefly provide some observations considering the covariance matrix \eqref{e:conf}, that is $\text{Cov}(\alpha^*) = \left(J(\alpha)^\top J(\alpha)\right)^{-1}$. In the case of one parameter, this covariance matrix is a scalar function
\begin{align*}
	\text{Cov}(\alpha^*) = \tfrac{1}{\|J(\alpha^*)\|_2^2}
\end{align*}
and the Jacobian is given by
\begin{align*}
	J(\alpha) =\begin{pmatrix}
		\tfrac{\partial}{\partial \alpha}\left(- C_\text{vol}(\alpha)x_0\right)\\
		\tfrac{\partial}{\partial \alpha}\left(- C_\text{vol}(\alpha)A\left(x_0 + B(\alpha)u_0\right)\right)\\
		\ldots\\
		\tfrac{\partial}{\partial \alpha}\left(- C_\text{vol}(\alpha) \left(A^{N-1} x_0 + \sum_{i=0}^{N-2} A^{N-1-i} B(\alpha)u_i\right)\right)
	\end{pmatrix}\in \mathbb{R}^N.
\end{align*}
where the $j$-th entry, $ j \in \{1,\ldots N-1\}$ is given by
\begin{align*}
	&\tfrac{\partial}{\partial \alpha}\left(- C_\text{vol}(\alpha)\left(A^jx_0 +  \sum_{i=0}^{j-1} A^{j-i}B(\alpha)u_i\right)\right) \\&= -C'(\alpha)\left(A^jx_0 +  \sum_{i=0}^{j-1} A^{j-1}B(\alpha)u_i\right) + \left(- C_\text{vol}(\alpha)\sum_{i=0}^{j-1} A^{j-i}B'(\alpha)u_i\right)
\end{align*}
% In this case,
% $$
% \mathcal{C}(\alpha)= \tfrac{1}{J(\alpha)^\top J(\alpha)}.
% $$
We provide some remarks considering this Jacobian.
\begin{itemize}
	\item Let $u_k=0$ for all $k$ and $x_0\neq 0$. If $(A,C'(\alpha^*))$ is observable and $N\geq \dim(x_0)$,  then $\|J(\alpha^*)\|^2_2 = \sum_{i = 0}^{N-1} |C'(\alpha^*) A^j x_0|^2 \neq 0$. A sufficient condition that the denominator does not vanish for $N<\dim(x_0)$ is that the observability index is at most $N$, where the observability index is the smallest $n_0$ such that
	$$[C'(\alpha^*),C'(\alpha^*)A,\ldots,C'(\alpha^*)A^{n_0-1}]$$ has full rank.
	\item If, on the other hand, $x_0=0$ and $u_k \neq 0$ for all $k$, then \begin{align*} C'(\alpha^*) \sum_{i=0}^{j-1} A^{j-i}B(\alpha^*)u_i = C'(\alpha^*)\left(B(\alpha^*),AB(\alpha^*),\ldots,A^{j-1}B(\alpha^*)\right) \left(\begin{smallmatrix}
			u_j\\
			:\\
			u_0
		\end{smallmatrix}\right)=0\end{align*} implies that $\left(\begin{smallmatrix}
	B(\alpha^*)u_j\\
	:\\
	B(\alpha^*)u_0
	\end{smallmatrix}\right)\in \ker \left(C'(\alpha^*),C'(\alpha^*)A,\ldots,C'(\alpha^*)A^{j-1}\right)$. If $j \geq \dim{x_0}$ and as $u_k \neq 0$ for all $k$, this can only hold if $(A,C'(\alpha^*))$ is not observable. Analogously one can argue for the term that contains $C$ and $B'$.
	%\item If we add a regularization term $\lambda\|\alpha-\alpha_d\|^2$ with a Tikhonov regularization parameter $\lambda > 0$ to the cost functional, we actually minimize the $2$-norm of the vector valued function
	%\begin{align*}
	%F_\text{reg}(\alpha) = \begin{pmatrix}
	%F(\alpha)\\
	%\sqrt\lambda (\alpha-\alpha_d)
	%\end{pmatrix}.
	%\end{align*}
	%In particular, the Jacobian is given by $ J_\text{reg}(\alpha) =\Delta F_\text{reg}(\alpha) = \begin{pmatrix}
	%J(\alpha)\\
	%\sqrt{\lambda}
	%\end{pmatrix}$. This means that $J_\text{reg}(\alpha)^\top J_\text{reg}(\alpha) = \lambda + J(\alpha)^\top J(\alpha)>0$. 
	%\item The comments in this part refer to the observability of $(A,C)$. The matrix $A=(I-\delta t A_c)^{-1}$ is an approximation of the matrix exponential for small time steps $\delta t$:
	%\begin{align*}
	%A = (I-\delta t A_c)^{-1} &= \sum_{k=0}^{\infty} \left( \delta t A_c\right)^k = I + \delta t A_c + \delta t^2A_c^2+\ldots\\
	%\exp(hA_c) &=  \sum_{k=0}^{\infty} \frac{\left( \delta t A_c\right)^k}{k!} = I + \delta t A_c + \frac{\delta t^2A_c^2}{2}+\ldots \\
	%\end{align*}
	\item Let $x_0 = 0$ and $u_k \equiv u = const.$ Then, the $j+1$-th entry of $J(\alpha)$ is given by
	\begin{align*}
		J(\alpha)_{j+1} = -C'_\text{vol}(\alpha)\left(A^jx_0 +  \sum_{i=0}^{j-1} A^{j-1}B(\alpha)u_i\right) + \left(- C_\text{vol}(\alpha)\sum_{i=0}^{j-1} A^{j-i}B'(\alpha)u_i\right) \\
		= -u\left(C'_\text{vol}(\alpha) \sum_{i=0}^{j-1} A^{j-i}B(\alpha) + C_\text{vol}(\alpha)\sum_{i=0}^{j-1} A^{j-i}B'(\alpha)\right).
	\end{align*}
	In particular, $\|J(\alpha_*)\|$ is strictly increasing in $u$ and hence, the width of the confidence intervals is inversely proportional to the norm of $u$. This also carries over to the two-parameter case, considering the entries of $J(\alpha) \in \mathbb{R}^{N\times 2}$.
\end{itemize}

\end{document}